\documentclass[preprint,12pt]{elsarticle}

\usepackage{kurz}
\usepackage{amssymb}
\usepackage{mathrsfs}
\usepackage{multirow}    
\usepackage{amsmath}
\usepackage{caption}
\usepackage{subcaption}
\usepackage{a4wide}
\usepackage{hyperref}
\hypersetup{hidelinks,
	colorlinks=true,
	allcolors=black,
	pdfstartview=Fit,
	breaklinks=true}
\usepackage{algorithm}
\usepackage{algpseudocode}
\usepackage{threeparttable}
\usepackage{color}

\usepackage{lineno}

\journal{Computer Methods in Applied Mechanics and Engineering}

\begin{document}
	
\begin{frontmatter}



\title{Transfer learning based physics-informed neural networks for solving inverse problems in engineering structures under different loading scenarios}


\author[ISM-BO]{Chen Xu}
\author[ISM-BO]{Ba Trung Cao}
\author[HPC]{Yong Yuan}
\author[ISM-BO]{Günther Meschke\corref{cor}\fnref{fncor}}

\cortext[cor]{Corresponding author}
\fntext[fncor]{email: guenther.meschke@rub.de}

\address[ISM-BO]{Institute for Structural Mechanics, Ruhr-University Bochum, Universitätsstraße 150, 44801 Bochum, Germany}
\address[HPC]{Department of Geotechnical Engineering, College of Civil Engineering, Tongji University, 1239 Siping Road, 20092 Shanghai, China}

\begin{abstract}
Recently, a class of machine learning methods called physics-informed neural networks (PINNs) has been proposed and gained prevalence in solving various scientific computing problems. This approach enables the solution of partial differential equations (PDEs) via embedding physical laws into the loss function. Many inverse problems can be tackled by simply combining the data from real life scenarios with existing PINN algorithms. In this paper, we present a multi-task learning method using uncertainty weighting to improve the training efficiency and accuracy of PINNs for inverse problems in linear elasticity and hyperelasticity. Furthermore, we demonstrate an application of PINNs to a practical inverse problem in structural analysis: prediction of external loads of diverse engineering structures based on limited displacement monitoring points. To this end, we first determine a simplified loading scenario at the offline stage. By setting unknown boundary conditions as learnable parameters, PINNs can predict the external loads with the support of measured data. When it comes to the online stage in real engineering projects, transfer learning is employed to fine-tune the pre-trained model from offline stage. Our results show that, even with noisy gappy data, satisfactory results can still be obtained from the PINN model due to the dual regularization of physics laws and prior knowledge, which exhibits better robustness compared to traditional analysis methods. Our approach is capable of bridging the gap between various structures with geometric scaling and under different loading scenarios, and the convergence of training is also greatly accelerated through not only the layer freezing but also the multi-task weight inheritance from pre-trained models, thus making it possible to be applied as surrogate models in actual engineering projects.
\end{abstract}

\begin{keyword}
Physics-informed neural networks (PINNs)
\sep Multi-task learning
\sep Transfer learning
\sep Inverse analysis
\sep Tunnel engineering
\end{keyword}

\end{frontmatter}




\section{Introduction}
\label{sec:Intro}
Machine learning has continued its forward movement over the past few years with advances in many exciting research areas, especially the branch of deep learning \cite{Lecun2015}, which is attributed to massive accumulation of data, advanced algorithms and availability of high performance computing. Meanwhile, an increasing number of applications of machine learning has been witnessed in the field of computational mechanics and engineering, e.g. fluid mechanics \cite{Duraisamy2019, Brunton2020, ML_acc_cfd}, molecular dynamics \cite{Fernandez2020, Zhang2018, Han2018, jacs_md}, learning constitutive laws \cite{Huang2020, Wang2019a, Linka2021, long2019pde}, solving fracture problems \cite{Hsu2020, Liu2020}, material property prediction \cite{Mozaffar2019, Li2019, Lu2020}, surrogate modeling \cite{PAPADOPOULOS2018411, Freitag2018, Cao2020, Cao:22}, model order reduction \cite{Cao_Freitag_Meschke:16, Koeppe2020, Liu2016} and multi-scale analysis \cite{Wang2018, Bessa2017, Saha2021, Logarzo2021}, to name a few examples. However, in most of these methods, a large amount of data collected from laboratory experiments or high-fidelity numerical simulations is inevitably required to establish a database for supervised learning. 
Neural networks (NNs) can then be treated as a "black box" to accomplish the nonlinear mapping between inputs and outputs. The robustness of NN highly depends on the noise level of experimental data, the accuracy of numerical simulations and the volume of data.

Recently, a different class of NNs known as physics-informed neural networks was introduced \cite{Raissi2019, Karniadakis} and has already gained a lot of attention in the scientific machine learning community. It is noted that the original idea of using neural networks to solve differential equations dates back to the end of the last century \cite{Lagaris1998, rico1993}, but due to the lack of computational power (computer storage and processing speed), this approach did not receive enough attention at that time. As a promising alternative of numerical discretization schemes, PINNs can solve partial differential equations (PDEs) by embedding the physical description (e.g. laws of physics and constraints) into NNs, without the requirement for big data as mentioned above. PINNs can leverage current domain knowledge and integrate seamlessly observational data and mathematical models, which are effective and efficient for both forward and inverse problems. A series of PINNs have been proposed by Karniadakis's research team, such as P-PINN for time-dependent PDEs \cite{Meng2020}, C-PINNs for domain decomposition \cite{Jagtap2020a}, N-PINNs for non-local models \cite{Pang2020}, V-PINNs with hp-refinement for function approximation \cite{Kharazmi2021}, and B-PINNs for uncertainty quantification~\cite{Yang2021, LINKA2022115346}. The original multilayer perceptron (MLP) in PINNs can be adjusted flexibly to include various network architectures, e.g. convolutional neural networks (CNNs) \cite{Gao2021}, long short-term memory (LSTM) networks \cite{Zhang2020}, generative adversarial networks (GANs) \cite{Yang2020}, and graph convolutional networks (GCNs) \cite{Gao2021a}. Currently, PINNs have been successfully applied to a wide range of scientific problems, e.g. fluid mechanics \cite{Raissi2020, Wessels2020, Sun2020, Jin2021, Wang2021, KASHEFI2022111510}, additive manufacturing \cite{Niaki2020, Zhu2021}, material defects identification \cite{Zhangenrui2022}, cardiovascular flows modeling \cite{Kissas2020}, photonic metamaterials \cite{Chenyuyao:20}, and reservoir simulation \cite{Gasmi2021}. Several open source libraries such as DeepXDE \cite{Lu2021}, SimNet \cite{Hennigh2021} and SciANN \cite{Haghighat2021a} were also developed to bridge the gap between PINN theory and its practical application. It is worth noting that PINN models sometimes suffer from training failures (e.g. slow convergence rate and bad minima), some remedies were thus proposed to resolve these difficulties, e.g. adaptive activation function \cite{Jagtap2020}, importance sampling approaches \cite{Nabian2021}, hard boundary imposition \cite{SUKUMAR2022114333, CHEN2021110624}, and gradient-enhanced methods \cite{Yu2022, CHIU2022114909}. A comprehensive review of the literature on PINNs is provided in \cite{cuomopinnreview}.

The emergence of physics-informed machine learning has brought new hopes to addressing solid mechanics problems in a different way. Haghighat et al. \cite{Haghighat2021} firstly presented the application of PINNs to forward and inverse analysis in solid mechanics, specifically linear elasticity and elasto-plasticity. Moreover, they developed a nonlocal PINN approach using the Peridynamic Differential Operator (PDDO) to better capture the solution with sharp gradients \cite{Haghighat2021b}. Inspired by the variational formulation of PDEs, Samaniego et al. \cite{Samaniego2020a} proposed the deep energy method (DEM) and rewrote the loss function of PINN as the potential energy, which opens a door to its application to various mechanical problems by defining the corresponding energies \cite{Nguyen-Thanh2020}. A more comprehensive comparison of governing equation-based (original PINNs) and energy-based (DEM) approaches can be referred to in \cite{Li2021}. Inspired by classical finite element methods (FEMs), Moseley et al. \cite{Moseley2021} proposed a new domain decomposition approach for PINNs called Finite Basis PINNs (FBPINNs), which are effective in solving problems with large domains and multi-scale solutions. Henkes et al. \cite{HENKES2022114790} applied C-PINNs to the heterogeneous microstructures and accurately resolved the localized strong nonlinear solution fields invoked by material inhomogeneities. Rezaei et al. \cite{REZAEI2022115616} proposed a mixed formulation-based PINN method using the spatial gradient of the primary variable as an output from separated neural networks, which has the potential to solve the unknowns in a heterogeneous domain without any available initial data from other sources.

Most of the existing papers focus on the application of PINNs to various forward problems, i.e. PINN is treated as a meshless numerical solver for partial differential equations. Despite that switching from numerical discretization schemes to mathematical optimization is promising, in terms of computational cost and accuracy, PINN still cannot outperform traditional computational methods such as FEMs. Therefore, in this paper, we exploit one of the best features of PINN, i.e. integration of observational data and physics models, to solve the inverse problem in structural engineering. Since the unbalanced back-propagated gradients during network training will pose a great challenge to the convergence of PINNs \cite{Wang2020}, some multi-task optimization methods are employed to mitigate the balance conflicts among different loss terms. We mention some of them below.
\begin{itemize}	
	\item [$\bullet$]\textbf{Mitigating Gradient Pathologies} \cite{Wang2020}: A learning rate annealing algorithm was presented to balance the interplay between different loss terms. During the training process, some instantaneous values are obtained by computing the ratio between the maximum gradient of PDE residual loss and the mean gradient of each loss term. The task weights are updated with these values using a moving average. 
	\item [$\bullet$]\textbf{GradNorm} \cite{gradnorm, HAGHIGHAT2022115141}: A gradient normalization algorithm was established to automatically balance learning tasks. All the losses can be trained at similar rates by dynamically tuning gradient magnitudes. 
	\item [$\bullet$]\textbf{Gradient Surgery (PCGrad)} \cite{yu2020gradient, WaiChingSun_multi:21}: Different loss terms may have conflicting gradient directions, which can lead to destructive interference. By projecting a task’s gradient onto the normal plane of the gradient of any other task that has a conflicting gradient, such detrimental interference can be greatly alleviated.
	\item [$\bullet$]\textbf{Multi-Objective Optimization} \cite{NEURIPS2018_Ozan}: The multi-task learning was formulated as multi-objective optimization. By optimizing the proposed multiple gradient descent algorithm upper bound (MGDA-UB), a Pareto optimal solution can be found.
	\item [$\bullet$]\textbf{Neural Tangent Kernel (NTK)} \cite{Wang2020a, Wang2021b}: The NTK matrix of PINNs was derived to analyze the training dynamics. A novel gradient descent algorithm was presented, which utilizes the eigenvalues of the NTK to adaptively calibrate the convergence rate of the different loss components.
	\item [$\bullet$]\textbf{Self-Adaptive PINNs (SA-PINNs)} \cite{mcclenny2020self}: The trainable self-adaptation weights were applied to each training point individually, so the neural network can give higher penalty to the points in difficult sub-regions in the domain. This method is reminiscent of the multiplicative soft attention mechanisms used in computer vision.
\end{itemize}

Although all these methods show promising results in forward problems, their performance in inverse analyses is insufficiently investigated. Most of them require plenty of additional gradient calculations/operations during training process, increasing the computational complexity. 

In this paper, we focus on the PINN for two types of constitutive models: linear elasticity and hyperelasticity (specifically, incompressible Neo-Hookean material). We first non-dimensionalize the governing equations, and a multi-task learning method using homoscedastic uncertainty \cite{Cipolla2018} is then introduced to allocate relative weightings to different learning objectives, which enhances the accuracy and accelerates the convergence of PINN at low computational cost. For comparison, we also implement the aforementioned SA-PINNs \cite{mcclenny2020self}, in which the authors claimed that their method outperformed other state-of-the-art PINN algorithm. By setting the boundary conditions as learnable parameters, PINN can accurately predict the loads acting on structures if supported by a sufficient amount of monitoring data. Usually in actual engineering projects, inverse analyses of structural components/elements in different sizes are in high demand, and the cost of obtaining monitoring data is really high. Therefore, a transfer learning based PINN approach is presented to fine-tune the network parameters (including network weights, biases, and task weights) on the basis of the pre-trained model prepared at the offline stage. Sufficiently accurate prediction results can be obtained with scaled geometries, more complicated loading scenarios, less computational cost, and less requirements for measurements, which exhibits great potential in solving inverse problems in structural analysis.

The remainder of this paper is organized as follows: A transfer learning based PINN method is developed in Section~\ref{sec:Methodology}. Numerical results for three benchmark examples in structural mechanics are shown in Section~\ref{sec:Benchmark}. Numerical results showcasing applications of the proposed approach to inverse analysis in engineering structures under different loading scenarios are presented in Section~\ref{sec:modeltransfer}. Finally, concluding remarks are given in Section~\ref{sec:Conc}.

\section{Methodology}
\label{sec:Methodology}
\subsection{PINNs: Physics-informed neural networks}
The general form of differential equations can be expressed as: 
\begin{equation}
	\begin{aligned}
		\mathcal{D}\left[ \boldsymbol{u}\left( \boldsymbol{X} \right) \right] &=h(\boldsymbol{X}), \boldsymbol{X}\in \varOmega,
		\\
		\mathcal{B}\left[ \boldsymbol{u}\left( \boldsymbol{X} \right) \right] &=g(\boldsymbol{X}), \boldsymbol{X}\in \partial \varOmega.
	\end{aligned}
\label{eq:pde}
\end{equation}
Here, $\mathcal{D}\left[ \cdot \right]$ denotes a differential operator acting on a function $\boldsymbol{u}$ (i.e. the solution to the differential equation), $\mathcal{B}\left[ \cdot \right]$ are boundary operators, $\boldsymbol{X}$ is an input vector in the domain $\varOmega$, and $\partial \varOmega$ stands for the boundary of $\varOmega$. This general formulation can be extended to foundational equations for a wide spectrum of engineering problems, such as thermodynamics, fluid dynamics, and elasticity.

Let us consider a fully connected feed-forward neural network $NN\left(\boldsymbol{X} ; \theta \right)$ with input vector $\boldsymbol{X}$, output vector $\boldsymbol{u}$, and network parameters $\theta$. This $L$-layer neural network, composed of an input layer, an output layer, and $L-2$ hidden layers, can approximate the solution function $\boldsymbol{u}\left(\boldsymbol{X}\right)$ through the following function composition: 
\begin{equation}
	\begin{aligned}
		\boldsymbol{u}\left( \boldsymbol{X} \right) \approx NN\left(\boldsymbol{X} ; \theta \right) &=f_L\circ \phi \circ f_{L-1}\circ \phi \circ f_{L-2}\circ \phi \cdots \phi \circ f_1\left( \boldsymbol{X} \right), \\
		f_l\left( \boldsymbol{a} \right) &=\boldsymbol{W}_l\cdot \boldsymbol{a}+\boldsymbol{b}_l, \quad l=1,\cdots ,L, \\
	\end{aligned}
\end{equation}
where $\phi$ is called activation function and directly determines the nonlinear property of the network. $f_l$ is a transformation function acting on the vector $\boldsymbol{a}$. The notation $\circ$ denotes function composition. $\boldsymbol{W}_l, \boldsymbol{b}_l$ are parameters between layers, known as weights and biases respectively, which constitute the set of trainable parameters $\theta$.  

\begin{figure}[!t]
\centering
\includegraphics{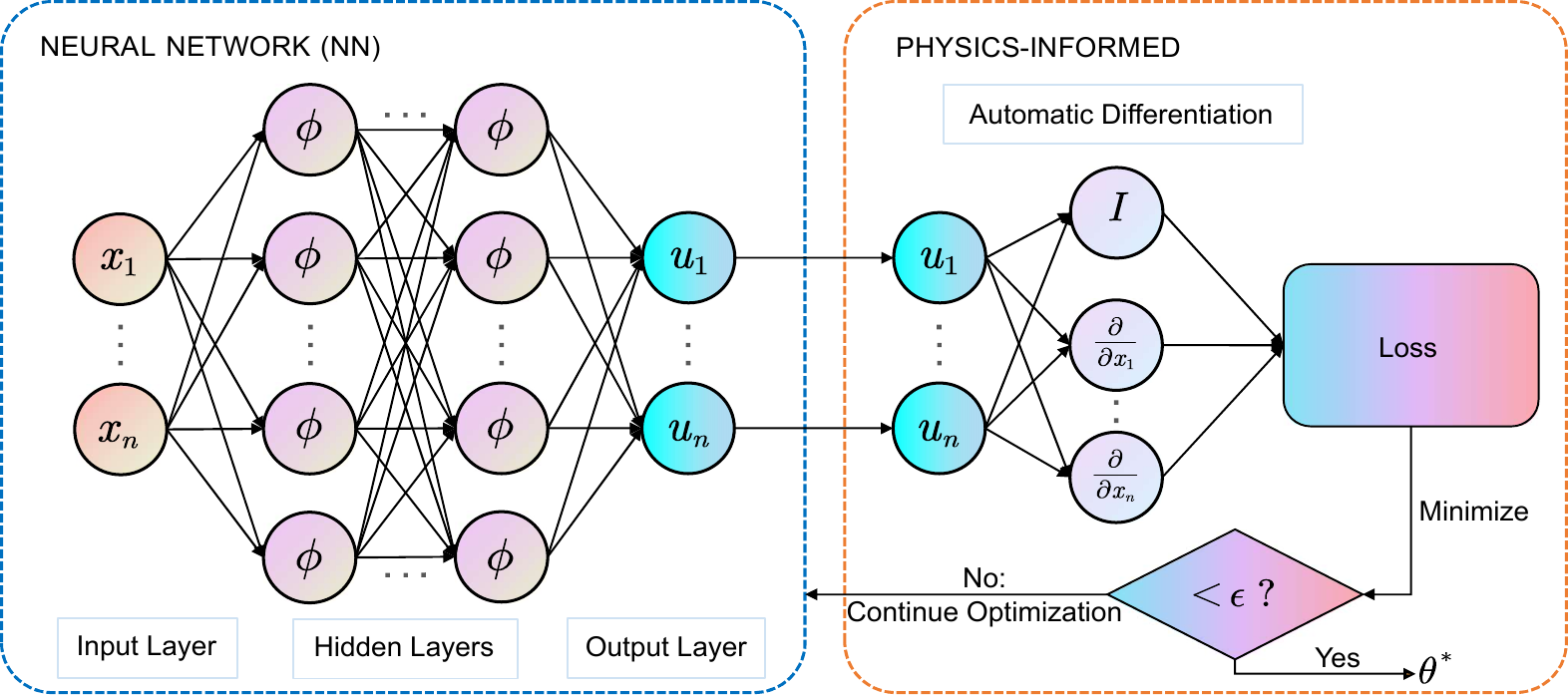}
\caption{Flowchart of a physics-informed neural network (PINN) framework}
\label{fig:flowchart}
\end{figure}

Given the architecture of a NN, a loss function should be chosen to measure the difference between prediction results obtained from NN and the actual values. In PINN models, loss functions are selected in the following form based on the definition in Eq.~\eqref{eq:pde}:
\begin{equation}
	\begin{aligned}
		\mathcal{L}(\theta)&=\mathcal{L}_{PDEs}(\theta)+\mathcal{L}_{BCs}(\theta), \\
		\mathcal{L}_{PDEs}(\theta)&=\frac{1}{N_{p}} \sum_{i}^{N_{p}}\left\|\mathcal{D}\left[NN\left(\boldsymbol{X}_i ; \theta\right) \right] -h(\boldsymbol{X}_i)  \right\|^{2}, \\
		\mathcal{L}_{BCs}(\theta)&=\frac{1}{N_{bc}} \sum_{j}^{N_{bc}}\left\|\mathcal{B}\left[NN\left(\boldsymbol{X}_j ; \theta\right) \right] -g(\boldsymbol{X}_j) \right\|^{2},
	\end{aligned}
\end{equation}
here $\mathcal{L}_{PDEs}(\theta)$ indicates the loss related to the PDEs, i.e. the embedded physical description, $\mathcal{L}_{BCs}(\theta)$ represents the loss related to the boundary conditions, $\left\{\boldsymbol{X}_i\right\}$ is a set of training points sampled over the whole domain $\varOmega~(i=1,2,\cdots,N_p$) and $\left\{\boldsymbol{X}_j\right\}$ is a set of training points sampled from the domain boundary $\partial \varOmega~(j=1,2,\cdots,N_{bc})$. Finally, the optimal network parameters $\theta ^*$ can be obtained by minimizing the loss function $\mathcal{L}(\theta)$. Note that while constructing the term $\mathcal{D}\left[NN\left(\boldsymbol{X}_i ; \theta\right) \right]$, the derivatives of outputs with respect to inputs can be obtained during back-propagation through the automatic differentiation (AD) module of Pytorch \cite{Paszke2019}. Based on the chain rule, AD can compute derivatives of arbitrary order automatically, accurately to working precision. Overall, the complete process of a PINN method is depicted in Fig.~\ref{fig:flowchart}. 

\subsection{Elasticity with non-dimensionalization}
For linear elasticity, the following PDE expresses the momentum balance equation, the constitutive relationship and the kinematic relations under small deformation assumption:
\begin{equation}
	\rho \frac{\partial^{2} \boldsymbol{u}}{\partial t^{2}}=\nabla((\lambda+\mu) \nabla \cdot \boldsymbol{u})+\nabla \cdot(\mu \nabla \boldsymbol{u})+\rho \boldsymbol{f}_{\boldsymbol{b}},
\end{equation}
where $\boldsymbol{u}$ denotes the displacement vector, $\rho$ represents the density of the material, $\lambda$ and $\mu$ are the Lamé parameters, and $\boldsymbol{f}_{\boldsymbol{b}}$ is a body force. In this paper, all the models involve static or quasi-static deformations where the acceleration term $\rho \frac{\partial^{2} \boldsymbol{u}}{\partial t^{2}}$ is neglected, and the body force $\boldsymbol{f}_{\boldsymbol{b}}$ is also absent. In that case,
\begin{equation}
	0=\nabla((\lambda+\mu) \nabla \cdot \boldsymbol{u})+\nabla \cdot(\mu \nabla \boldsymbol{u}).
\end{equation}

We introduce the following dimensionless variables with scaling factors $u_c$ and $l_c$ \cite{Langtangen2016}: 
\begin{equation}
	\bar{\boldsymbol{u}}=\frac{\boldsymbol{u}}{u_c}, \quad \bar{x}=\frac{x}{l_c}, \quad \bar{y}=\frac{y}{l_c}, \quad \bar{z}=\frac{z}{l_c}.
\end{equation}

Inserting the dimensionless variables gives:
\begin{equation}
	0=l_c^{-2} u_{c} \bar{\nabla}\left(\left(\lambda+\mu\right) \bar{\nabla} \cdot \bar{\boldsymbol{u}}\right)+l_c^{-2} u_{c} \mu \bar{\nabla} \cdot(\bar{\nabla} \bar{\boldsymbol{u}}).
\end{equation}

Dividing by $l_c^{-2} u_{c} \mu$ leads to:
\begin{equation}
	0=\bar{\nabla}((\frac{\lambda}{\mu} + 1) \bar{\nabla} \cdot \bar{\boldsymbol{u}})+\bar{\nabla} \cdot(\bar{\nabla} \bar{\boldsymbol{u}}).
\end{equation}

The Cauchy stress tensor $\boldsymbol{\sigma}$ is computed by:
\begin{equation}
	\boldsymbol{\sigma}=\lambda tr(\boldsymbol{\varepsilon}) \boldsymbol{I}+ 2\mu \boldsymbol{\varepsilon}=\lambda \nabla \cdot \boldsymbol{u} \boldsymbol{I}+\mu\left(\nabla \boldsymbol{u}+(\nabla \boldsymbol{u})^{T}\right).
	\label{eq:stress_tensor}
\end{equation}

With the dimensionless variables, Eq.~\eqref{eq:stress_tensor} becomes:
\begin{equation}
	\boldsymbol{\sigma}=\lambda u_{c} l_c^{-1}  \bar{\nabla} \cdot \bar{\boldsymbol{u}}+\mu u_{c} l_c^{-1} \left(\bar{\nabla} \bar{\boldsymbol{u}}+(\bar{\nabla} \bar{\boldsymbol{u}})^{T}\right).
\end{equation}

By defining the scaling factor $\sigma_{c}$ as:
\begin{equation}
	\sigma_{c}=\mu u_{c} l_c^{-1},
\end{equation}
we then can finally get the dimensionless stress tensor:
\begin{equation}
	\bar{\boldsymbol{\sigma}}=\frac{\boldsymbol{\sigma}}{\sigma_{c}}=\frac{\lambda}{\mu} \bar{\nabla} \cdot \bar{\boldsymbol{u}}+\left(\bar{\nabla} \bar{\boldsymbol{u}}+(\bar{\nabla} \bar{\boldsymbol{u}})^{T}\right).
\end{equation}

\subsection{Hyperelasticity with non-dimensionalization}
Let us consider a solid made of a homogeneous, isotropic, incompressible Neo-Hookean material. The vector $\boldsymbol{X}$ is used to define the undeformed reference configuration, and $\boldsymbol{u}$ denotes the displacement vector. The deformation gradient $\boldsymbol{F}$ is given by:
\begin{equation}
	\boldsymbol{F}=\boldsymbol{I}+\frac{\partial \boldsymbol{u}}{\partial \boldsymbol{X}},
	\label{eq:deform_grad}
\end{equation}
where $\boldsymbol{I}$ is the identity tensor, and Eq.~\eqref{eq:deform_grad} is kinematics. 

To ensure the incompressibility of a hyperelastic material, i.e. $J:=\det ( \boldsymbol{F} ) =1$, the strain energy model can be written in the form:
\begin{equation}
	\varPsi =\frac{\mu}{2}( I_1-3 )-p_{hydro}(J-1),
\end{equation}
in which $\mu$ is the second Lamé parameter, the first strain invariant is defined as $I_1=\mathrm{trace}(\boldsymbol{C})$, and $p_{hydro}$ is hydrostatic pressure which functions as a Lagrangian multiplier. $\boldsymbol{C}$ refers to the right Cauthy-Green tensor and is defined by $\boldsymbol{C}=\boldsymbol{F}^T\cdot \boldsymbol{F}$. 

The first Piola-Kirchhoff stress $\boldsymbol{\sigma}^{PK1}$ is calculated by
\begin{equation}
	\boldsymbol{\sigma}^{PK1}=\frac{\partial \varPsi}{\partial \boldsymbol{F}}=-p_{hydro}\boldsymbol{F}^{-T}+\mu \boldsymbol{F}.
	\label{eq:pk1}
\end{equation}

We introduce the scaling factor $\sigma_{c}=\mu$, then Eq.~\eqref{eq:pk1} becomes:
\begin{equation}
	\boldsymbol{\bar{\sigma}}^{PK1}=\frac{\boldsymbol{\sigma}^{PK1}}{\sigma_c}=-\frac{p_{hydro}}{\mu}\boldsymbol{F}^{-T}+\boldsymbol{F}=-\bar{p}_{hydro}\boldsymbol{F}^{-T}+\boldsymbol{F}.
\end{equation}

The final equilibrium PDE is expressed by:
\begin{equation}
	\bar{\nabla}_{\boldsymbol{\bar{X}}} \cdot \boldsymbol{\bar{\boldsymbol{\sigma}}}^{PK1}=0,
\end{equation}
where $\bar{\nabla}_{\boldsymbol{\bar{X}}} \cdot \boldsymbol{\bar{\boldsymbol{\sigma}}}^{PK1}$ denotes the divergence operator applying on the scaled first Piola-Kirchhoff stress w.r.t dimensionless $\boldsymbol{\bar{X}}$ on the initial configuration.

For clarity, the dimensionless variables and equilibrium equations involved in our PINN model for two materials are given in the following Table~\ref{tab:constitutive}.
\begin{table}[!t]
	\renewcommand{\arraystretch}{1.5}
	\centering
	\caption{Dimensionless equations for elasticity/hyperelasticity}
	\begin{tabular}{lcc}
		\hline
		& Linear elasticity & Hyperelasticity \\ \hline
		Dimensionless variables & \multicolumn{2}{c}{$\bar{\boldsymbol{X}}=\boldsymbol{X}/{l_c},\, \bar{\boldsymbol{u}}=\boldsymbol{u}/u_c$}                                \\
		Scaling factor $\sigma_{c}$ & $\sigma_{c}=\mu u_{c} l_c^{-1}$        & $\sigma_{c}=\mu$        \\
		Stress tensors  & $\bar{\boldsymbol{\sigma}}=\lambda/\mu \cdot \bar{\nabla} \cdot \bar{\boldsymbol{u}}+\left(\bar{\nabla} \bar{\boldsymbol{u}}+(\bar{\nabla} \bar{\boldsymbol{u}})^{T}\right)$                 & $\boldsymbol{\bar{\sigma}}^{PK1}=-\bar{p}_{hydro}\boldsymbol{F}^{-T}+\boldsymbol{F}$                                \\
		Equilibrium     & $\bar{\nabla}_{\boldsymbol{\bar{X}}} \cdot \boldsymbol{\bar{\boldsymbol{\sigma}}}=0$                 & $\bar{\nabla}_{\boldsymbol{\bar{X}}} \cdot \boldsymbol{\bar{\boldsymbol{\sigma}}}^{PK1}=0$                                \\ \hline
	\end{tabular}
	\label{tab:constitutive}
\end{table}

\subsection{Multi-task learning using uncertainty}
As mentioned in Section \ref{sec:Intro}, we adopt the multi-task learning method in \cite{Cipolla2018} to weigh the losses for each individual task. It is reported that the homoscedastic uncertainty in Bayesian modeling can capture the relative confidence between tasks, which can thus be described as task-dependent uncertainty for weighting losses. We define the Gaussian likelihood $p$ with mean given by the network output:
\begin{equation}
	\begin{aligned}
		p\left(\boldsymbol{u} \mid NN(\boldsymbol{X};\theta )\right)=\mathcal{N}\left(NN(\boldsymbol{X};\theta ), \alpha^{2}\right),
	\end{aligned}
\end{equation}
with the scalar $\alpha$ as a noise parameter. In a setting of a neural network with $K$ tasks, the multi-objective likelihood can be written as:
\begin{equation}
	\begin{aligned}
		p\left(\boldsymbol{u}_{1}, \ldots, \boldsymbol{u}_{K} \mid NN(\boldsymbol{X};\theta )\right)=p\left(\boldsymbol{u}_{1} \mid NN(\boldsymbol{X};\theta )\right) \ldots p\left(\boldsymbol{u}_{K} \mid NN(\boldsymbol{X};\theta )\right).
	\end{aligned}
\end{equation}

In maximum likelihood estimation (MLE), it is often convenient to work with the natural logarithm of the likelihood function, and we can therefore maximize: 

\begin{equation}
	\begin{aligned}
		\log p\left(\boldsymbol{u}_{1}, \ldots, \boldsymbol{u}_{K} \mid NN(\boldsymbol{X};\theta )\right)
		&=\log p\left(\boldsymbol{u}_{1} \mid NN(\boldsymbol{X};\theta )\right) + \ldots + \log p\left(\boldsymbol{u}_{K} \mid NN(\boldsymbol{X};\theta )\right) \\
		&=\log \mathcal{N}\left(\boldsymbol{u}_{1} ; NN(\boldsymbol{X};\theta ), \alpha_{1}^{2}\right) + \ldots + \log \mathcal{N}\left(\boldsymbol{u}_{K} ; NN(\boldsymbol{X};\theta ), \alpha_{K}^{2}\right) \\
		&= (-\frac{1}{2\alpha_{1}^{2}} \mathcal{L}_{1}  - \log \alpha_{1}) + \ldots + (-\frac{1}{2\alpha_{K}^{2}} \mathcal{L}_{K}  - \log \alpha_{K}) \\
		&= (-\frac{1}{2\alpha_{1}^{2}} \mathcal{L}_{1}  - \frac{1}{2}\log \alpha_{1}^{2}) + \ldots + (-\frac{1}{2\alpha_{K}^{2}} \mathcal{L}_{K}  - \frac{1}{2}\log \alpha_{K}^{2}) \\
		&\propto (-\frac{1}{2\alpha_{1}^{2}} \mathcal{L}_{1}  - \log (1+\alpha_{1}^{2})) + \ldots + (-\frac{1}{2\alpha_{K}^{2}} \mathcal{L}_{K}  - \log (1+\alpha_{K}^{2})).  
	\end{aligned}
\label{eq:likelihood}
\end{equation}

Here we explain the last line of Eq.~\eqref{eq:likelihood}. Because the logarithmic function $\log \alpha^{2}$ takes negative values when $\alpha^{2}$ lies in the interval of $(0,1)$, we rewrite it as $\log (1+\alpha^{2})$ to avoid the loss of becoming negative during training \cite{liebel2018auxiliary}.

This leads to our final minimisation target: 
\begin{equation}
	\begin{aligned}
	    \mathcal{L}(\theta, \alpha_{1}, \ldots, \alpha_{K})=\sum_{i}^{K} (\frac{1}{2\alpha_{i}^{2}}\mathcal{L}_{i} + \log (1+\alpha_{i}^{2})),
	\end{aligned}
\label{eq:eq21}
\end{equation}
where $\mathcal{L}_{i}$ is the loss related to task $i$, and the trainable parameters are also extended to include the noise parameter (i.e. task weight) $\alpha$. This approach is not only mathematically interpretable, but also intuitively reasonable. In some extreme cases, when the loss of a task is very large and the losses of other tasks are very small, the multi-objective optimization degenerates to single-objective optimization and the update of network parameters is dominated by the task with a larger loss. For the adopted approach (see Eq.~\eqref{eq:eq21}), if a task is easy to optimize (i.e., the uncertainty of the task is small), a smaller $\alpha$ can be learned so that the model does not neglect this simple task. Moreover, $\log (1+\alpha^{2})$ can be regarded as a regularization term which prevents the learned $\alpha$ from being too large.

\subsection{Proposed approach: Transfer learning based boundary-condition-learnable PINNs}
Let us start from the PINN model for linear elasticity. The scaled Cartesian coordinates $\{\bar{x}, \bar{y}, \bar{z}\}$ ($\{\bar{x}, \bar{y}\}$ for 2D problems) are considered as input layer while scaled displacements $\{\bar{u}_{x}, \bar{u}_{y}, \bar{u}_{z}\}$ ($\{\bar{u}_{x}, \bar{u}_{y}\}$ for 2D problems) are taken as output layer. Consider a set of collocation training points $\{ \boldsymbol{X}_{i}^{col} \} _{i=1}^{N_{col}}$, observational data points $\{ \boldsymbol{X}_{i}^{data} \} _{i=1}^{N_{data}}$, free boundary condition points $\{ \boldsymbol{X}_{i}^{freebc} \} _{i=1}^{N_{freebc}}$, Neumann boundary condition points $\{ \boldsymbol{X}_{i}^{nbc_{j}} \} _{i=1}^{N_{nbc_{j}}}$, Dirichlet boundary condition points $\{ \boldsymbol{X}_{i}^{dbc} \} _{i=1}^{N_{dbc}}$, where $N_{col}$, $N_{freebc}$, $N_{nbc_{j}}$, $N_{dbc}$ and $N_{data}$ denote the number of points, respectively. Generally, the PINN loss function is of the form:

\begin{equation}
	\begin{aligned}
	    \mathcal{L}(\boldsymbol{\Theta}) &= \mathcal{L}(\theta, \alpha_{0}, \alpha_{1}, \alpha_{2}, \alpha_{3}, \alpha_{4}, P_1, \ldots, P_{N_{load}}) = \sum_{i=0}^{4} (\frac{1}{2\alpha_{i}^{2}}\mathcal{L}_{i} + \log (1+\alpha_{i}^{2})), \\
	    \mathcal{L}_0 &= MSE_{dbc}=\frac{1}{N_{dbc}}\sum_{i=1}^{N_{dbc}}{\left\| \boldsymbol{\bar{u}}(\boldsymbol{X}_{i}^{dbc};\boldsymbol{\Theta })-\boldsymbol{\bar{u}}^*(\boldsymbol{X}_{i}^{dbc}) \right\| ^{2}}, \\
	    \mathcal{L}_1 &= MSE_{pde} = \frac{1}{N_{\mathrm{col}}}\sum_{i=1}^{N_{col}}{\left\| \bar{\nabla}_{\bar{\boldsymbol{X}}}\cdot \bar{\boldsymbol{\sigma }}(\boldsymbol{X}_{i}^{col};\boldsymbol{\Theta}) \right\| ^{2}}, \\
	    \mathcal{L}_2 &= MSE_{freebc} = \frac{1}{N_{freebc}}\sum_{i=1}^{N_{freebc}}{\left\| \bar{\boldsymbol{\sigma }}(\boldsymbol{X}_{i}^{freebc};\boldsymbol{\Theta })\cdot \boldsymbol{N}\left( \boldsymbol{X}_{i}^{freebc} \right) \right\| ^{2}}, \\
	    \mathcal{L}_3 &= \sum_{j=1}^{N_{load}}{MSE_{nbc_j}}=\sum_{j=1}^{N_{load}}{\frac{1}{N_{nbc_j}} \sum_{i=1}^{N_{nbc_j}}{\left\| \bar{\boldsymbol{\sigma }}(\boldsymbol{X}_{i}^{nbc_j};\boldsymbol{\Theta })\cdot \boldsymbol{N}\left( \boldsymbol{X}_{i}^{nbc_j} \right) -P_j \right\| ^{2}}}, \\
	    \mathcal{L}_4 &= MSE_{data}=\frac{1}{N_{data}}\sum_{i=1}^{N_{data}}{\left\| \bar{\boldsymbol{u}}(\boldsymbol{X}_{i}^{data};\boldsymbol{\Theta })-\bar{\boldsymbol{u}}^*(\boldsymbol{X}_{i}^{data}) \right\| ^{2}},
	\end{aligned}
	\label{eq:totloss}
\end{equation}
here, $\alpha_{i}$ is the noise parameter for each task and $\boldsymbol{N}$ is the unit normal vector to the boundary surface. The dimensionless stress tensor $\bar{\boldsymbol{\sigma}}$ is obtained from the first order derivative of the outputs $\bar{\boldsymbol{u}}$ w.r.t the inputs $\bar{\boldsymbol{X}}$. 

\begin{figure}[!t]
\centering
\includegraphics[scale=0.69]{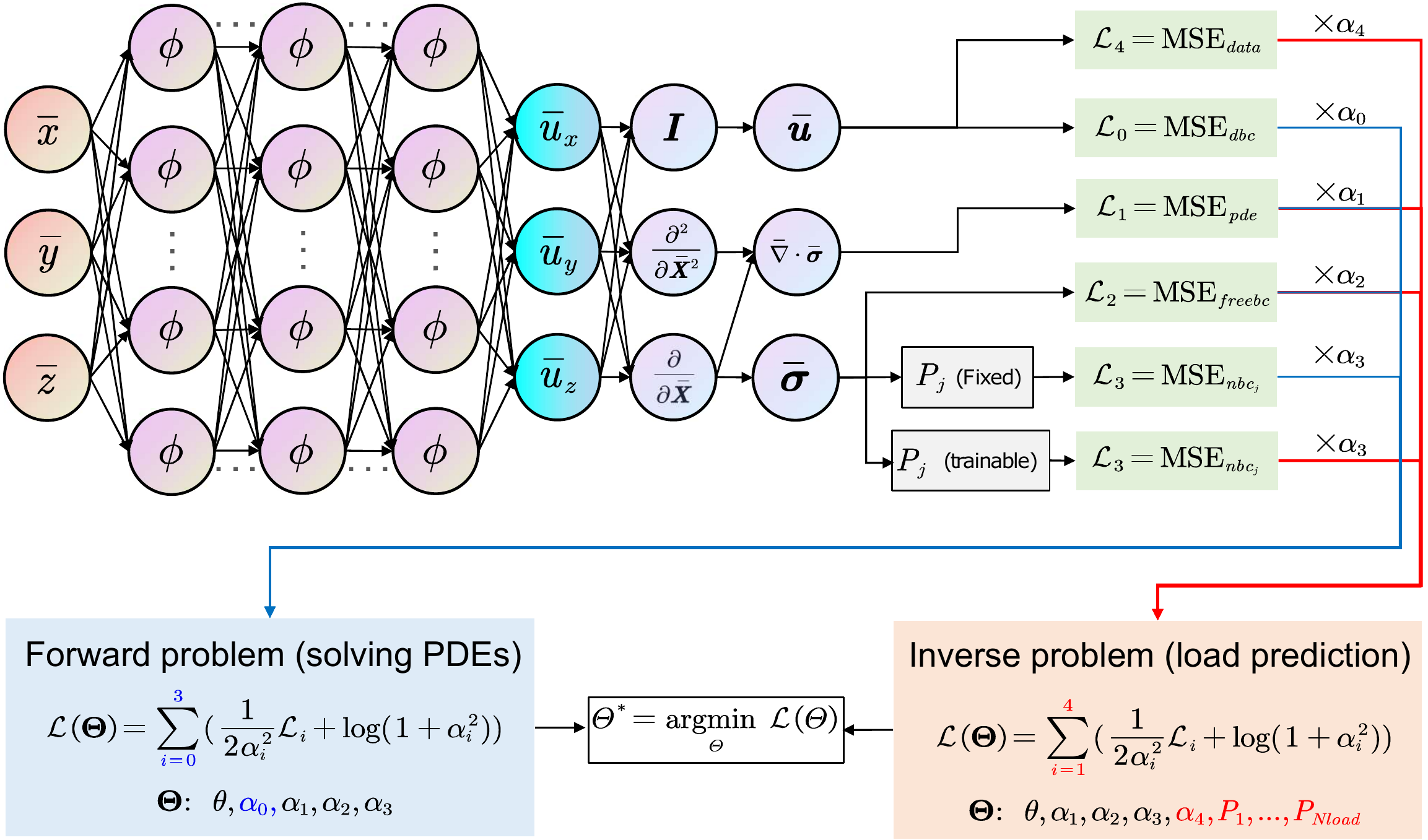}
\caption{A demonstration of the setup of the proposed PINN method for linear elasticity (with different learnable parameters marked in blue and red respectively)}
\label{fig:nnsetup}
\end{figure} 

As shown in Fig.~\ref{fig:nnsetup}, the PINN model for elasticity can be trained for two different purposes by customizing the loss function $\mathcal{L}(\boldsymbol{\Theta})$ with the set of various loss tasks:
\begin{itemize}
	\item [$\bullet$]\textbf{Forward problems}: Specifically for solving linear elastic problems using PINNs, architectures can be modified to satisfy exactly the governing PDEs ($\mathcal{L}_1$), free boundary conditions ($\mathcal{L}_2$), Neumann boundary conditions ($\mathcal{L}_3$), and Dirichlet boundary conditions ($\mathcal{L}_0$). The training parameters $\boldsymbol{\Theta}$ are hereby comprised of network parameters $\theta$ and task weights $\{\alpha_0, \alpha_1, \alpha_2, \alpha_3\}$. 
	\item [$\bullet$]\textbf{Inverse problems}: When an inverse analysis (e.g. load prediction) is required, the loss term $\mathcal{L}_4$ can be further incorporated into $\mathcal{L}(\boldsymbol{\Theta})$. $\boldsymbol{u}^*$ are observational data measurements of displacement $\boldsymbol{u}$ which can be collected from either in-situ tests or from laboratory experiments. Note that the loss $\mathcal{L}_0$ is discarded here, because the observational data can automatically help fulfill the prescription of Dirichlet boundary conditions. In addition to the general trainable parameters $\{\theta, \alpha_1, \alpha_2, \alpha_3, \alpha_4\}$ which are similar to that in forward problems, the unknown physical quantities $\{P_j\}_{j=1}^{N_{load}}$ are also involved, where $N_{load}$ denote the number of loads. 
\end{itemize}

\begin{figure}[!t]
	\centering
	\includegraphics[scale=0.69]{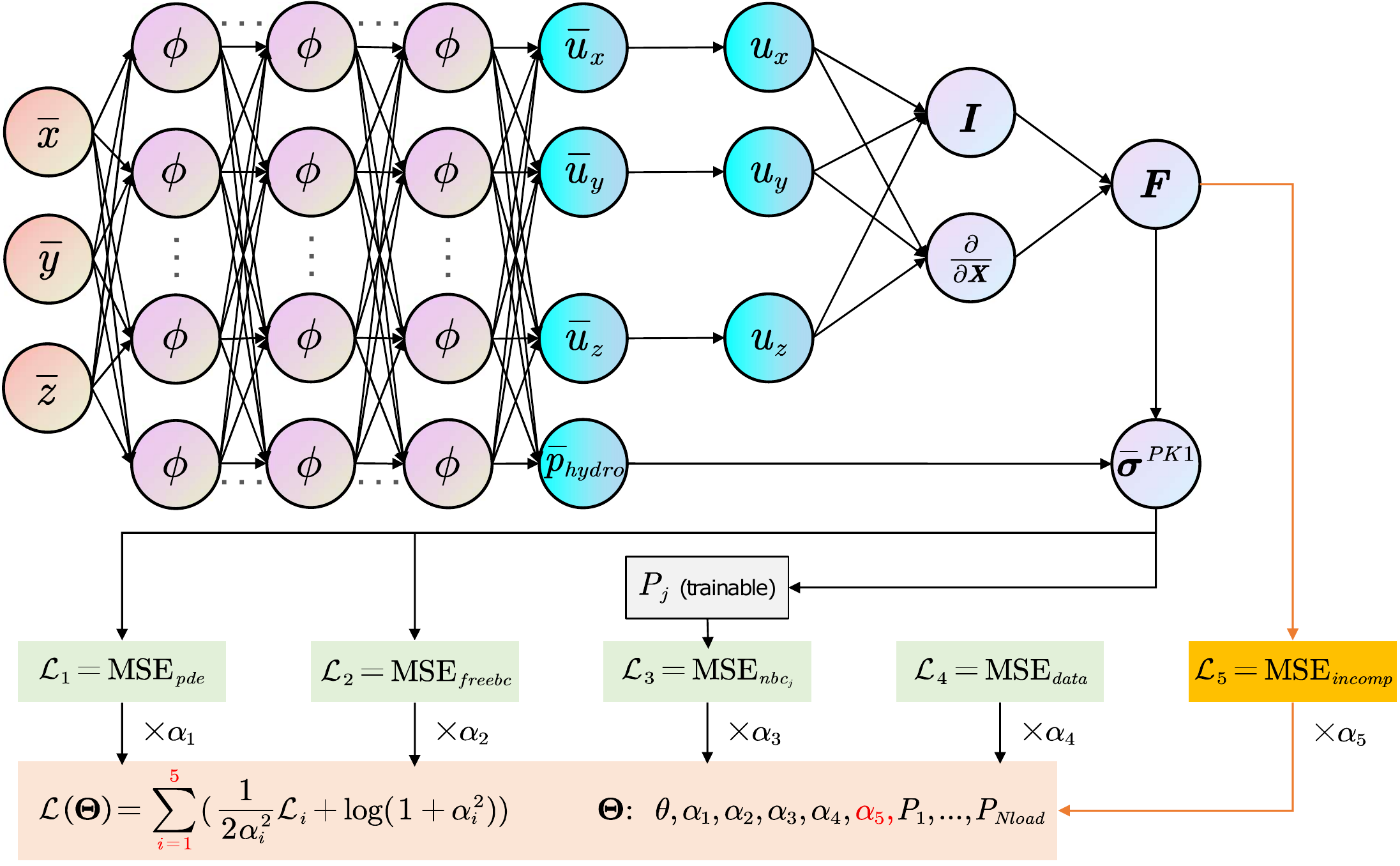}
	\caption{The architecture of PINNs for load prediction in incompressible hyperelastic materials}
	\label{fig:hyperPINN}
\end{figure}

Next, we extend our PINN framework from linear elasticity to hyperelasticity. As shown in Fig.~\ref{fig:hyperPINN}, herein we only present the PINN architecture for inverse analyses in incompressible Neo-Hookean materials. The input of the neural network remains the same while the output is augmented with the scaled hydrostatic pressure $\bar{p}_{hydro}$. The total loss $\mathcal{L}(\boldsymbol{\Theta})$ consists of five terms, of which $\mathcal{L}_1$, $\mathcal{L}_2$, $\mathcal{L}_3$, and $\mathcal{L}_4$ are formally consistent with Eq.~\eqref{eq:totloss}, while the scaled Cauchy stress tensor $\bar{\boldsymbol{\sigma}}$ needs to be replaced with the scaled first Piola-Kirchhoff stress tensor $\boldsymbol{\bar{\sigma}}^{PK1}$. Furthermore, the new physical constraint, i.e. the incompressibility condition, is satisfied by:
\begin{equation}
	\begin{aligned}
		\mathcal{L}_5=MSE_{incomp}=\frac{1}{N_{col}}\sum_{i=1}^{N_{col}}{\left\| \det \left( \boldsymbol{F}(\boldsymbol{X}_{i}^{col};\boldsymbol{\Theta }) \right) -1 \right\| ^2}.
	\end{aligned}
\end{equation}

\begin{figure}[!t]
	\centering
	\includegraphics{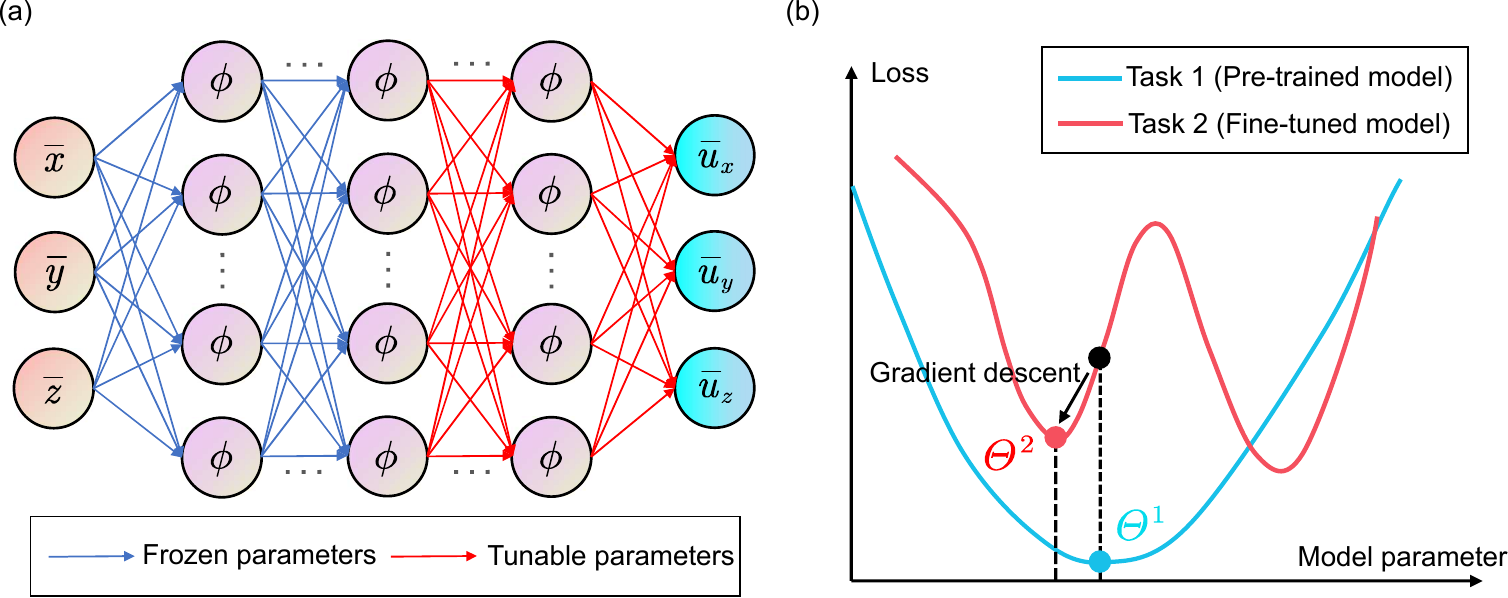}
	\caption{A schematic of transfer learning: (a) update of network parameters and (b) loss function convergence over iteration steps}
	\label{fig:transfer}
\end{figure}

Engineering structures are mostly composed of geometrically similar solid instances, for example, tunnels are assembled with identical segmented lining rings. This enlightened us to adopt transfer learning \cite{pan2009survey}, which is a very popular approach in the field of natural language processing \cite{LEEHY01, LEEHY02, LEEHY03} and computer vision. Several previous studies combining PINN and transfer learning have been carried out for solving forward problems \cite{Niaki2020, Goswami2020, Chakraborty2021}. Conventional machine learning is designed to work in isolation for specific tasks. Models have to be rebuilt from ground up once the scenario changes. In transfer learning, knowledge from previously trained models can be leveraged for new tasks, which can significantly accelerate the training process. As depicted in Fig.~\ref{fig:transfer}~(a), in the transfer learning process, weights and biases of some layers (the initial three layers in this paper) from the previously trained neural network will be frozen, while the remaining parameters will be fine-tuned with a smaller learning rate. The advantage of the approach is three-folds. First, frozen parameters ensure that the extracted features are retained, which allows the geometric similarity between structures to be exploited \cite{Goswami2020}. Second, as shown in Fig.~\ref{fig:transfer}~(b), the blue curve represents the loss of the pre-trained model for task 1 with respect to the model parameters. The optimal network parameters $\boldsymbol{\Theta}^1$ can be obtained by minimizing the loss function. Due to the similarity between task 1 and task 2, the optimal parameters $\boldsymbol{\Theta}^1$ is very likely to be very close to the optimal parameters $\boldsymbol{\Theta}^2$ of the fine-tuned model for the new task 2. Therefore, convergence can be achieved with much fewer epochs. Third, the saved task weights allow each loss term to be minimized in a balanced way at the beginning of the retraining stage (further demonstrated in Section~\ref{sec:tunnel}). As a result, the total number of parameters to be updated is greatly reduced and the number of training epochs required to get convergence is smaller. The overall framework of the proposed PINN for the inverse analysis of engineering structures is presented in Algorithm~\ref{alg:algo}.

In this paper, the neural networks in all cases are composed of 5 hidden layers with the hyperbolic tangent activation function (tanh) and each layer has 30 neurons. Synthetic data obtained from finite element simulations are considered as measured data. The machine learning processes were implemented in Pytorch \cite{Paszke2019} and trained in Google Colab on a NVIDIA Tesla P100 GPU. The parameters were optimized using the Adam optimizer \cite{Kingma2015}. Details on the hyperparameter settings are shown in Table~\ref{tab:hyperparam}. 

\begin{algorithm}[H]
\caption{Transfer learning based boundary-condition-learnable PINNs for the inverse analysis of loading conditions on engineering structures} \label{alg:algo}
\begin{algorithmic}[1]
\State Determine a simplified loading scenario (geometry, material property, and external loads, etc.) for structures to be analyzed at the offline stage.
\State Build a finite element model and run the simulation to obtain displacement results $\boldsymbol{u}^*(\boldsymbol{X}_{i}^{col})$ at all nodes of the structure.
\State Generate the coordinates of sampling points $\{\boldsymbol{X}_{i}^{col}\}$, $\{\boldsymbol{X}_{i}^{freebc}\}$, $\{\boldsymbol{X}_{i}^{nbc_{j}}\}$.
\State Choose appropriate scaling factors and non-dimensionalize all the physical quantities.
\State Initialize the neural network $NN(\boldsymbol{x};\theta)$, task weights $\alpha_i$ and unknown loads $P_j$.
\State Minimize the loss function to compute the optimal parameter $\boldsymbol{\Theta}^*_0$.   \Comment{Eq.~\eqref{eq:totloss}}
\State Save the neural network parameters $\theta^*_0$ and task weights $\alpha_i^*$.
\For{$n = 1,\dots,N$}
\State Select the structure $S_n$ to be analyzed at the online stage.
\State Obtain the displacements $\boldsymbol{u}^*(\boldsymbol{X}_{i}^{data})$ at some selected monitoring points $\{ \boldsymbol{X}_{i}^{data} \}$.
\State Generate the coordinates of sampling points $\{\boldsymbol{X}_{i}^{col}\}$, $\{ \boldsymbol{X}_{i}^{freebc}\}$, $\{ \boldsymbol{X}_{i}^{nbc_{j}}\}$.  
\State Choose appropriate scaling factors and non-dimensionalize the physical quantities.
\State Load the neural network parameters $\theta^*_0$ and task weights $\alpha_i^*$ as a pre-trained model.
\State Initialize the unknown loads $P_j$.
\State Freeze the initial three layers and fine-tune the remaining network parameters.
\State Minimize the loss function to compute the optimal parameter $\boldsymbol{\Theta}^*_n$. 
\Comment{Eq.~\eqref{eq:totloss}}
\State \textbf{\underline{Results}: predicted external loads $P_j$ acting on the structure $S_n$.}
\EndFor
\end{algorithmic}
\end{algorithm}

\begin{table}[H]
	\centering
	\caption{Hyperparameters setting in the codes}
	\begin{tabular}{lcc}
		\hline
		& Initial value                 & Learning rate \\ \hline
		Neural network parameter $\theta$       & Default weight initialization & 0.001         \\
		Task weights $\alpha_i$                 & 1.0                           & 0.0001        \\
		Unknown loads $P_j$                     & 1.0                           & 0.001         \\
		Tunable NN parameters in transfer learning & -                             & 0.0005        \\ \hline
	\end{tabular}
	\label{tab:hyperparam}
\end{table}

\section{Benchmark examples}
\label{sec:Benchmark}
In the following, we illustrate the performance of the proposed PINN method in the context of three benchmark examples. Section~\ref{sec:cylinder} demonstrates effectiveness of non-dimensionalization and multi-task learning through a classical 2D elastic problem \cite{Samaniego2020a}. Section~\ref{sec:plate22} focuses on a 2D hyperelastic problem with stress concentrations, which is tricky for plain PINNs \cite{FUHG2022110839}. Section~\ref{sec:3dbeam} presents the robustness of our PINN method to Gaussian noise in observational data through a 3D elastic problem. 

The idea of non-dimensionalization is to eliminate the units of each loss term, thus making the summation in the loss function more reasonable. Through further appropriate selection of scaling factors, the input and output of the neural network can be automatically normalized. In this paper, in order to realize Min-max normalization, the maximum values of geometrical and displacement quantities are directly selected as $l_c$ and $u_c$, respectively. Table~\ref{tab:scalefac} presents all scaling factors used in this paper for clarity. 

\begin{table}[!b]
	\centering
	\caption{The scaling factors for non-dimensionalization in all numerical examples}
		\begin{tabular}{lcc}
			\hline
			& $l_c$ & $u_c$ \\ 
			\hline
			Section~\ref{sec:cylinder} (thick cylinder) & 5.0 & 0.0003 \\
			Section~\ref{sec:plate22} (plate $2\times2$) & 1.0 & 0.03 \\
			Section~\ref{sec:3dbeam} (3D beam) & 5.0 & 0.04 \\
			Section~\ref{sec:plate4224} (plate $4\times2$) & 2.0 & 0.08 \\
			Section~\ref{sec:plate4224} (plate $2\times4$) & 2.0 & 0.047 \\
			Section~\ref{sec:tunnel} (tunnel case $2p$) &4.0 & 0.0018 \\
			Section~\ref{sec:tunnel} (tunnel case $4p\_a$) & 4.0 & 0.0024 \\
			Section~\ref{sec:tunnel} (tunnel case $4p\_b$) & 4.0 & 0.002 \\ \hline
		\end{tabular}
	\label{tab:scalefac}
\end{table}

\subsection{Prediction of internal pressure acting on a thick elastic cylinder}
\label{sec:cylinder}
\begin{figure}[!b]
	\centering
	\includegraphics{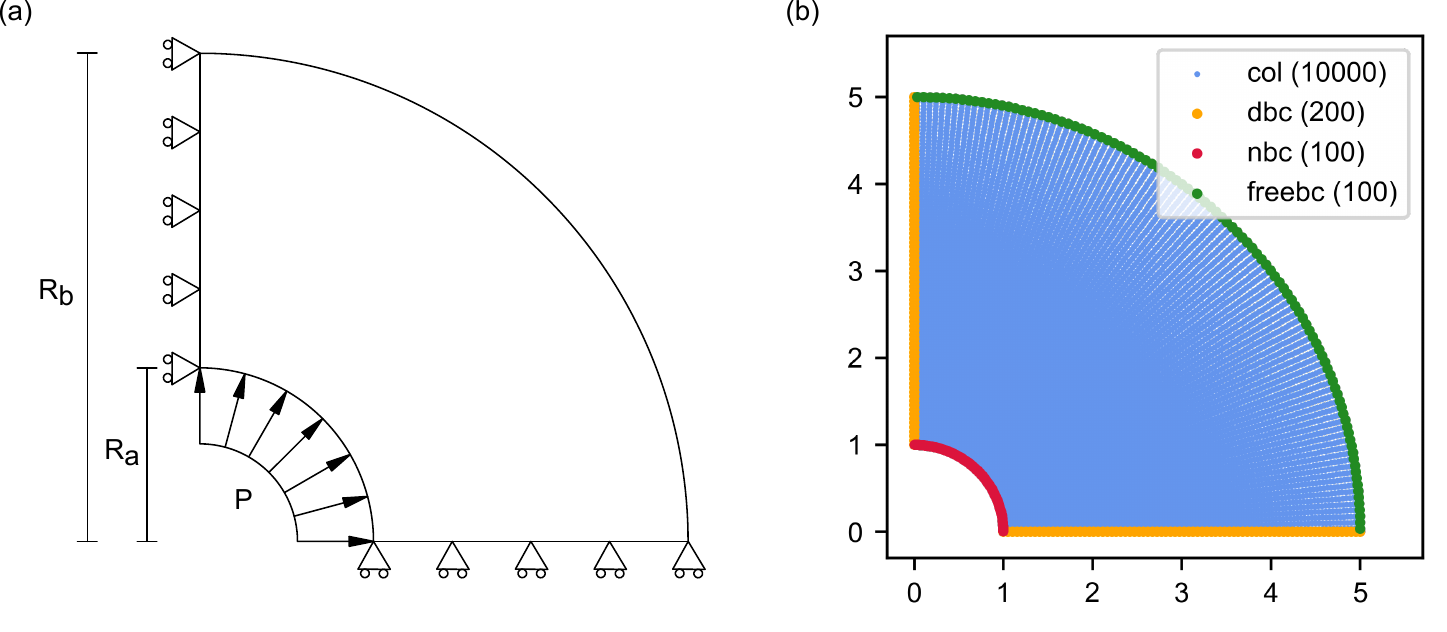}
	\caption{Thick cylinder under internal pressure: (a) geometry and loading condition and (b) sampled training points distribution (Numbers of points are indicated in parentheses.)}
	\label{fig:bench_geometry}
\end{figure}

The first example considers of a thick cylinder which is subjected to internal pressure. The analysis is formulated as a 2D plane stress problem. Four different methods, i.e. plain PINN (without non-dimensionalization), scaled PINN (with non-dimensionalization but without multi-task learning), SA-PINN (scaled) and PINN with uncertainty weighting (scaled), are implemented for comparison. For SA-PINNs in this paper, we initialize the self adaptive weights on the data points to be $U(0,1)$ and the learning rates for all self-adaptive weights are set to 0.001. Figure~\ref{fig:bench_geometry}~(a) illustrates the geometry and the boundary conditions for this problem. The neural networks are trained using a grid of $100\times100$ training points, i.e. $N_{col}$ = 10000, over the computational domain (see Fig.~\ref{fig:bench_geometry}~(b)). The boundary conditions are trained with an additional 100 training points per edge.

As a reference, the analytical solution for the displacement field in X and Y-axis is
\begin{equation}
	\begin{aligned}
	    u_{x}(r,\beta) &=\frac{R_{a}^{2} P r}{E\left(R_{b}^{2}-R_{a}^{2}\right)}\left(1-\nu+\left(\frac{R_{b}}{r}\right)^{2}(1+\nu)\right) \cos \beta, \\
        u_{y}(r,\beta) &=\frac{R_{a}^{2} P r}{E\left(R_{b}^{2}-R_{a}^{2}\right)}\left(1-\nu+\left(\frac{R_{b}}{r}\right)^{2}(1+\nu)\right) \sin \beta,
	\end{aligned}
\label{eq:analytical}
\end{equation}
and the components of the stress field obtained are provided below:
\begin{equation}
	\begin{aligned}
	    \sigma_{rr}(r,\beta) &=\frac{R_{a}^{2} P}{R_{b}^{2}-R_{a}^{2}}\left(1-\frac{R_{b}^{2}}{r^{2}}\right), \\
        \sigma_{\beta\beta}(r,\beta) &=\frac{R_{a}^{2} P}{R_{b}^{2}-R_{a}^{2}}\left(1+\frac{R_{b}^{2}}{r^{2}}\right), \\
        \sigma_{r \beta}(r,\beta) &=0,
	\end{aligned}
\end{equation}
where $\{r,\beta\}$ denote polar coordinates with origin in the center. The inner radius and outer radius of the cylinder are $R_{a}=1$ and $R_{b}=5$ respectively. The pressure $P$ is applied on the inner circular edge with $P=20$. The material properties considered for this example are $E=1\times10^{5}$ and $\nu=0.3$. The Dirichlet boundary conditions are a-priori fulfilled by:
\begin{equation}
	\begin{aligned}
	    \hat{u}_x=\bar{u}_x\times \bar{y}, \quad \hat{u}_y=\bar{u}_y\times \bar{x}.
	\end{aligned}
\end{equation}

\begin{figure}[!b]
	\centering
	\includegraphics{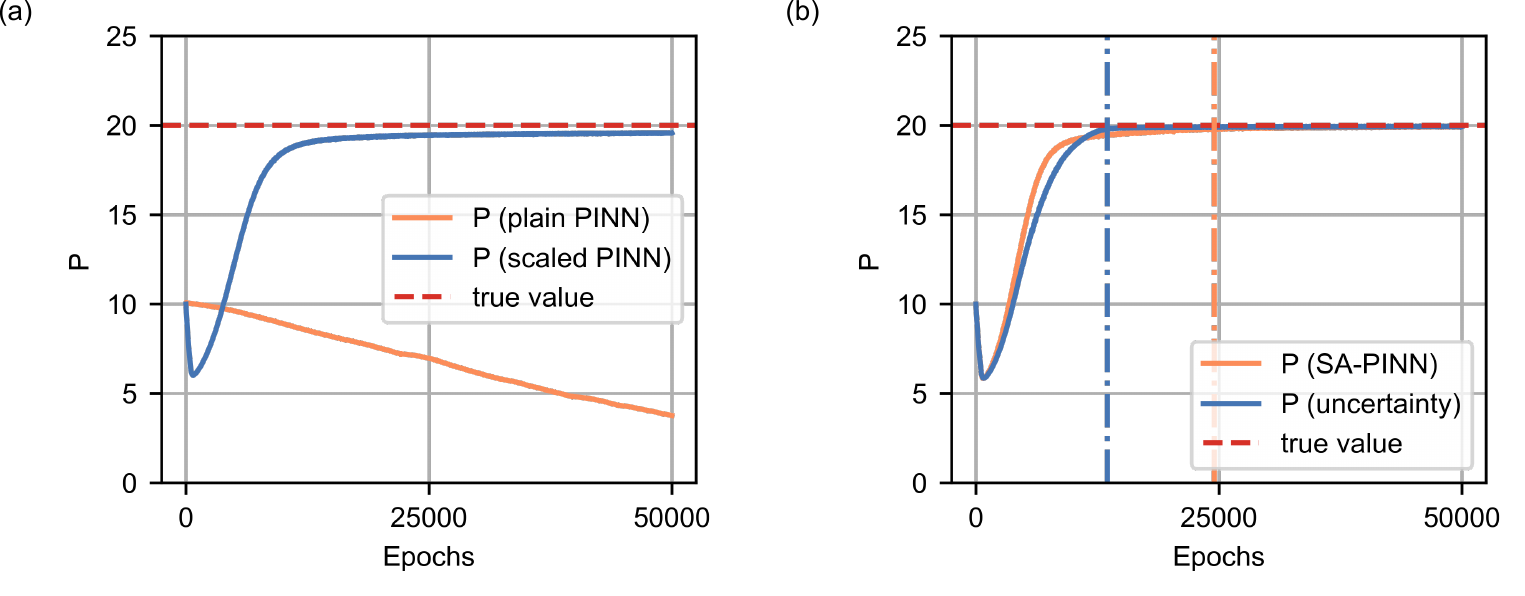}
	\caption{Results of loads for the thick cylinder problem: (a) plain PINN versus scaled PINN and (b) SA-PINN versus PINN with uncertainty weighting (The first epoch to converge to less than $1\%$ prediction error is marked with dash-dotted line.)}
	\label{fig:cylinder_load}
\end{figure}

\begin{figure} [!b]
	\centering
	\includegraphics[scale=0.93]{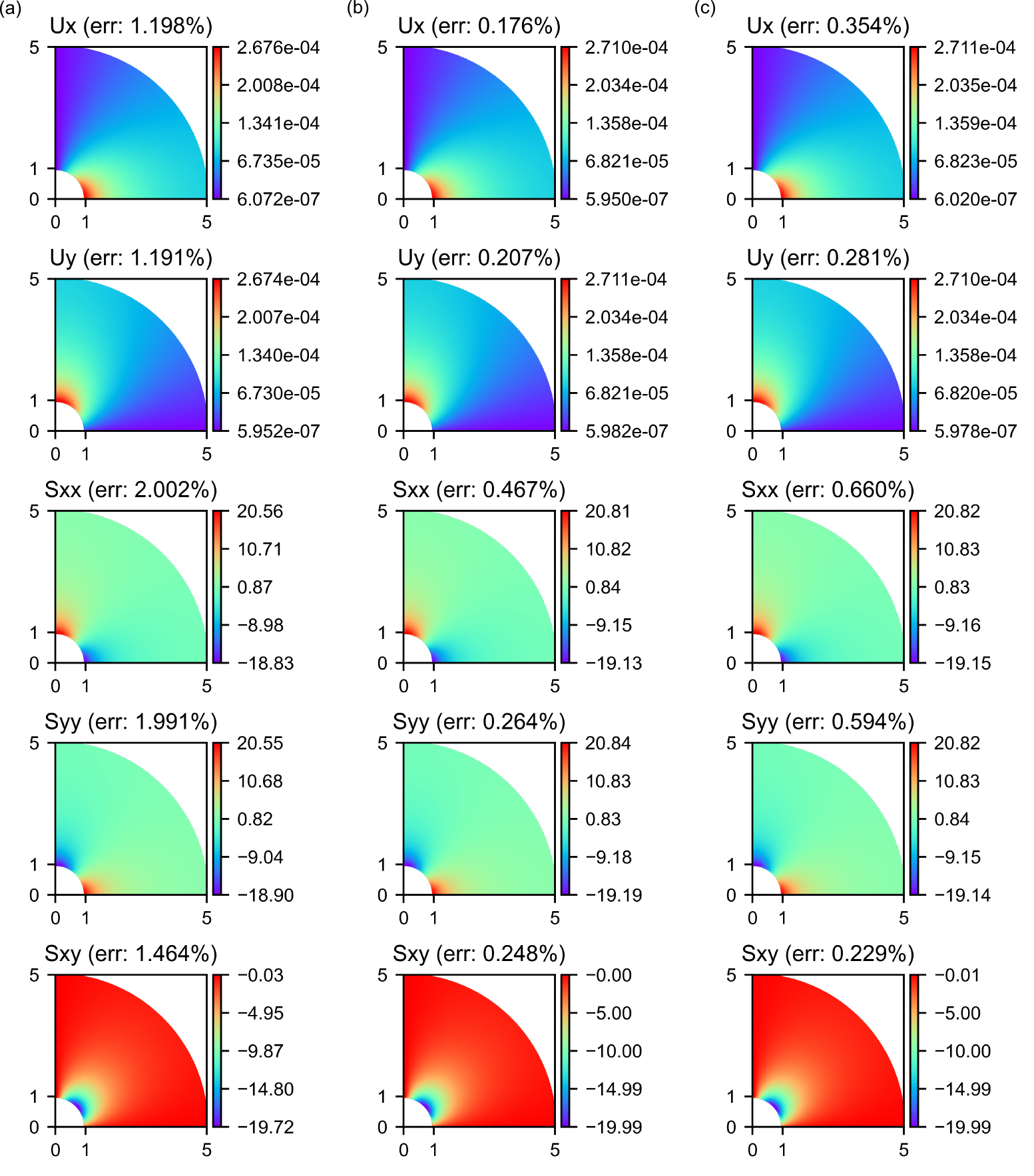}
	\caption{Displacement and stress components of thick cylinder under internal pressure: (a) scaled PINN, (b) SA-PINN and (c) PINN with uncertainty weighting}
	\label{fig:bench_field} 
\end{figure}

Here we set the acting load $P$ as an unknown parameter, and the target of this example is to identify the value of $P$ based on the analytical solution for the displacement field in Eq.~\ref{eq:analytical}. Figure~\ref{fig:cylinder_load} shows the predicted loads of the four methods after 50000 training steps. Plain PINN completely fails to predict the final load while the scaled PINN yields a much better result with $P=19.569$ (error: $2.155\%$), which indicates that the non-dimensionalization can significantly improve the performance of PINN. Both SA-PINN and PINN with uncertainty weighting can further increase the accuracy, $P=19.931$ (error: $0.345\%$) and $P=19.929$ (error: $0.355\%$), respectively. In Fig.~\ref{fig:bench_field} we compare the prediction results of field outputs. Compared to that of scaled PINN, the maximum $L_2$ error of field outputs of the other two multi-task learning approaches is reduced to less than $1\%$.

\begin{figure}[!b]
	\centering
	\includegraphics{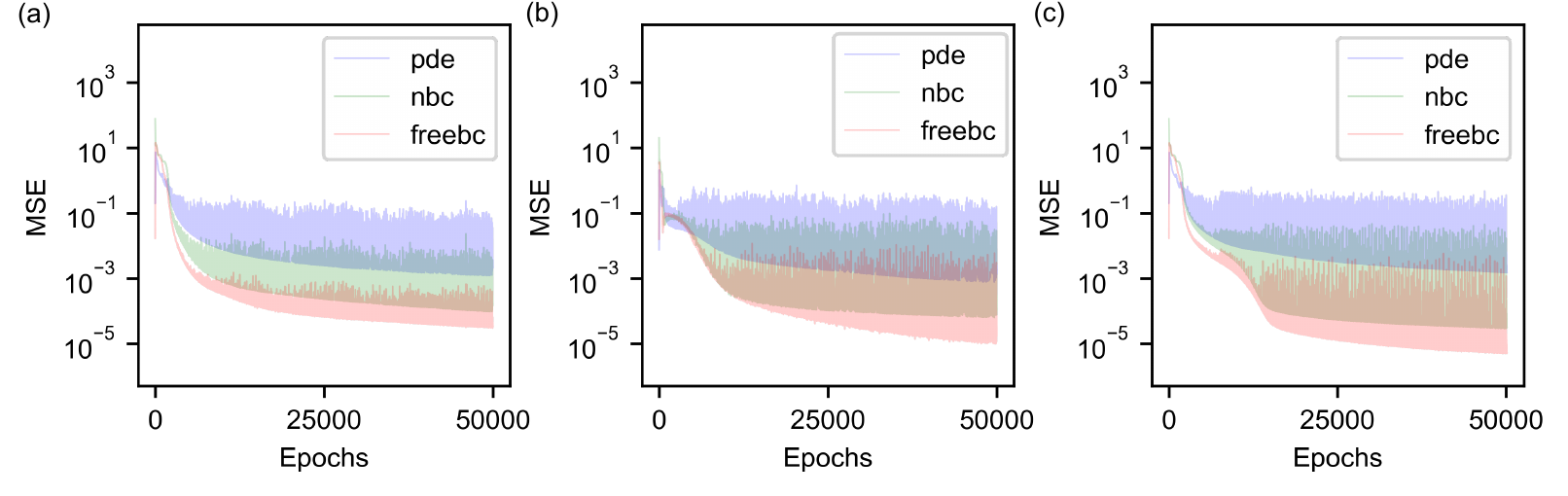}
	\caption{Evolution of loss terms for the thick cylinder problem: (a) scaled PINN, (b) SA-PINN and (c) PINN with uncertainty weighting}
	\label{fig:bench_loss}
\end{figure}

\begin{figure}[!b]
	\centering
	\includegraphics{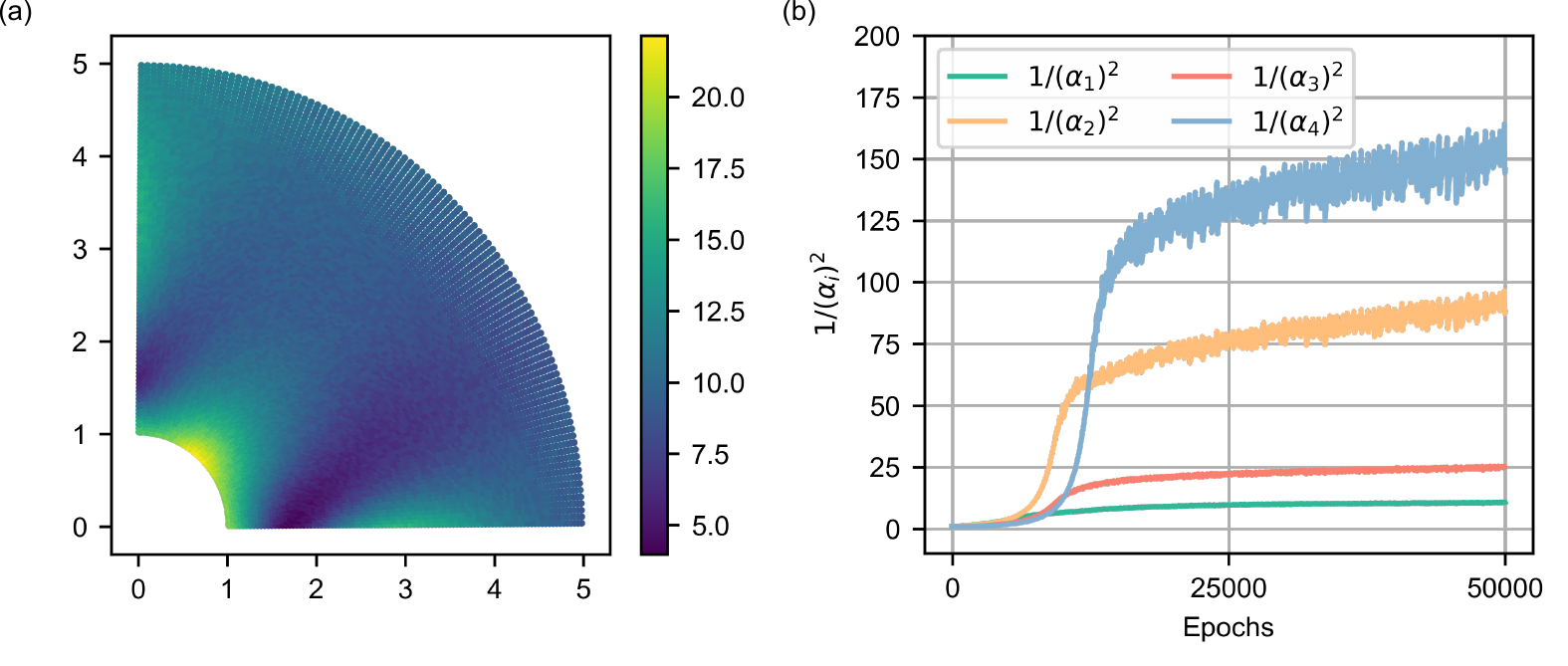}
	\caption{Results of multi-task weights for the thick cylinder problem in: (a) SA-PINN (Brighter colored points correspond to larger weights.) and (b) PINN with uncertainty weighting (We plot $1/(\alpha_i)^2$ here instead of the task weight $\alpha_i$ to better demonstrate the relative magnitude of task weights.)}
	\label{fig:bench1_weight}
\end{figure}

Figure~\ref{fig:bench_loss} shows the convergence of each individual term in the physics loss. The results of multi-task weights are given in Fig.~\ref{fig:bench1_weight}. Figure~\ref{fig:bench1_weight} (b) shows that $\mathcal{L}_4$ has the largest weight, indicating that the model has to pay extra attention to the data loss, resulting in an increase in accuracy. The final prediction of SA-PINN is slightly more accurate than PINN with uncertainty weighting, however, the latter converges to the result within $1\%$ error faster at epoch 13500 while SA-PINN obtains the equivalent error at epoch 24500 (see Fig.~\ref{fig:cylinder_load} (b)). Since pointwise self-adaptive weights are defined in the SA-PINN, these additional training parameters are proportional to the number of training points. In contrast, the additional parameters of our uncertainty weighting method depend only on the number of tasks, so the computational complexity is reduced, and results with acceptable accuracy can be obtained at low cost. 

\subsection{Prediction of line load acting on a square hyperelastic plate}
\label{sec:plate22}

\begin{figure}[!b]
	\centering
	\includegraphics{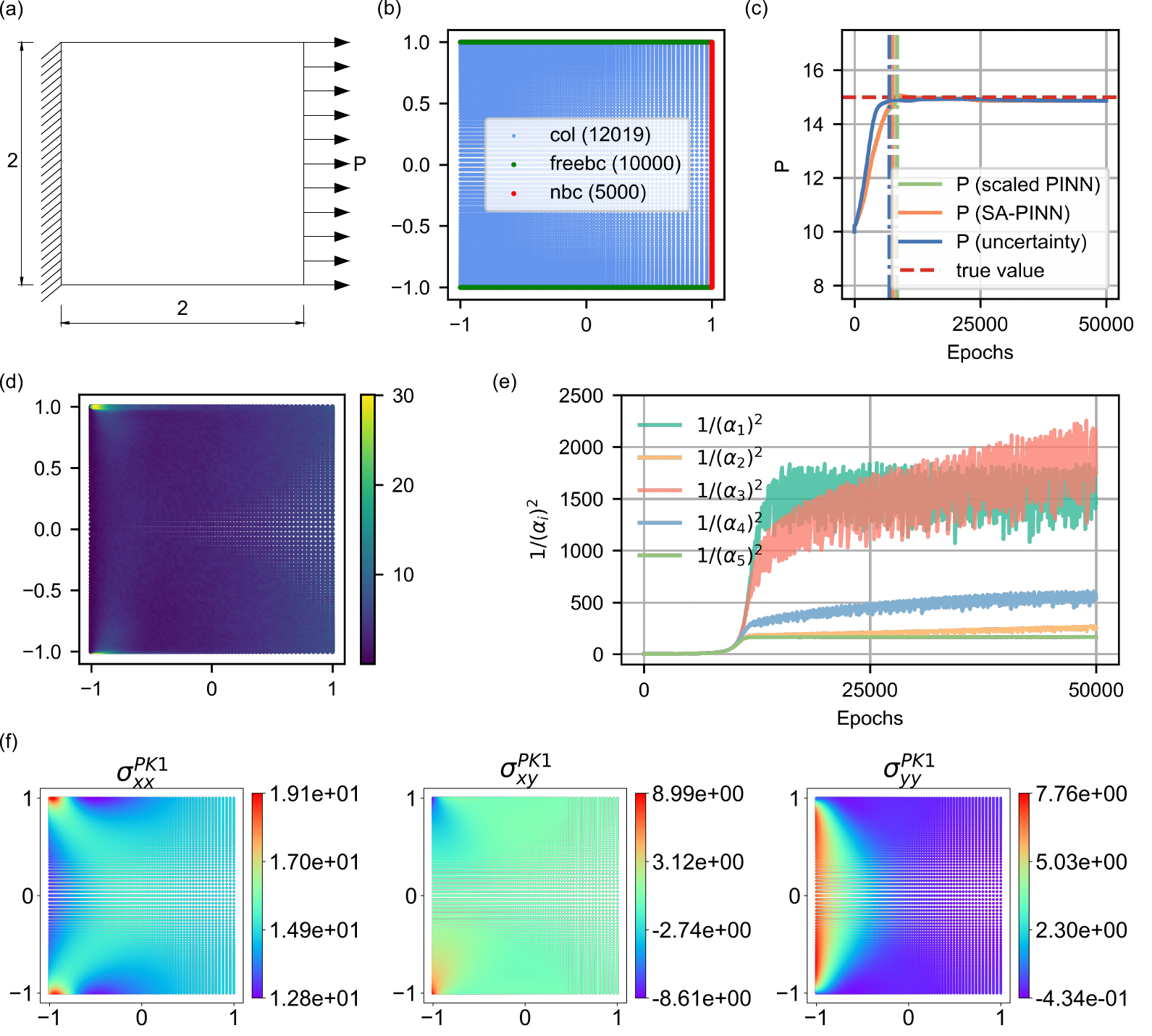}
	\caption{PINN model for prediction of the hyperelastic square plate subjected to a uniform line load: (a) geometry and loading condition, (b) sampled training points distribution (Numbers of points are indicated in parentheses.), (c) results of load prediction (The first epoch to converge to less than $1\%$ prediction error is marked with dash-dotted line.), (d) self-adaptive weights in SA-PINN (Brighter colored points correspond to larger weights.), (e) results of multi-task weights for PINN with uncertainty weighting and (f) first Piola-Kirchhoff stress components obtained from PINN with uncertainty weighting}
	\label{fig:plate22}
\end{figure}

In the second example we continue to study the performance of PINN with uncertainty weighting, and the model in this section will be saved for further transfer learning in Section~\ref{sec:plate4224}. A $2\times2$ square block made of incompressible Neo-Hookean materials is subjected to a uniform line load $P=20$ on the right-hand side while the left-hand side is clamped (see Fig.~\ref{fig:plate22}~(a)). The material properties of the plate are $E=1000$ and $\nu=0.5$. The FEM solution for reference is generated from a $100\times100$ elements mesh with grid refinements on the left. The collocation points are the same as the FEM nodes, and the boundary conditions points are selected uniformly with the spacing of 0.0004 (see Fig.~\ref{fig:plate22}~(b)). Similarly, the Dirichlet boundary conditions are enforced by defining:
\begin{equation}
	\begin{aligned}
		\hat{u}_x=\bar{u}_x\times (\bar{y}+1), \quad \hat{u}_y=\bar{u}_y\times (\bar{x}+1).
	\end{aligned}
\end{equation}

Considering the load $P$ as a learnable parameter, the goal of this example is to predict the value of $P$ based on the displacement field $\boldsymbol{u}^*(\boldsymbol{X}_{i}^{col})$ from FEM. Figure~\ref{fig:plate22}~(c) shows the results of the load after 50000 training epochs. The final predictions $P$ for scaled PINN, SA-PINN (scaled) and PINN with uncertainty weighting (scaled) are 14.847 (error: $1.020\%$), 14.886 (error: $0.760\%$) and 14.858 (error: $0.947\%$), respectively. The first epochs to converge to the result within $1\%$ error for the three methods are 8500, 7500 and 6900, respectively. Sufficiently accurate results are obtained in all three methods, while a faster convergence rate is observed in the PINN with uncertainty weighting. Figure~\ref{fig:plate22}~(e) depicts the multi-task weights obtained by our adopted method.
As shown in Fig.~\ref{fig:plate22}~(f), although the loading condition in this example is simple, stress concentrations occur at the left side of the domain as a result of the clamped constraint. Given that plain PINN can fail to follow the expected pattern \cite{FUHG2022110839}, this example again illustrates the performance of PINN with non-dimensionalization and multi-task learning.

\subsection{Prediction of tensile loads acting on a 3D elastic beam}
\label{sec:3dbeam}
\begin{figure}[!t]
	\centering
	\includegraphics{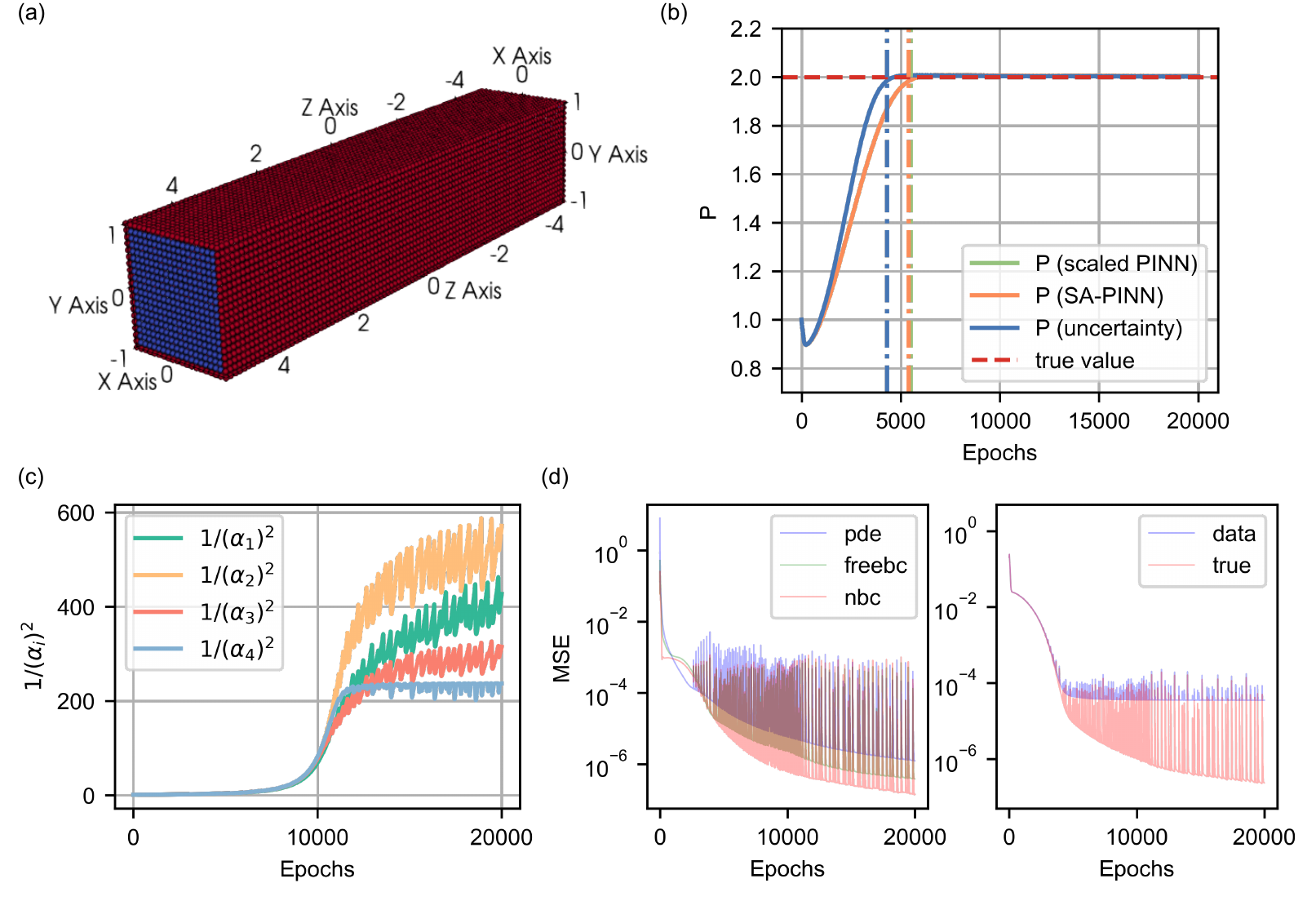}
	\caption{PINN model for prediction of the 3D elastic beam under uniaxial tension: (a) training point distribution, (b) results of load prediction (The first epoch to converge to less than $1\%$ prediction error is marked with dash-dotted line.), (c) results of multi-task weights for PINN with uncertainty weighting and (d) evolution of physics loss (left) and data loss (right)}
	\label{fig:3Dbeam}
\end{figure}

The third example considers a $10\times2\times2$ elastic beam as shown in Fig.~\ref{fig:3Dbeam}~(a). A tensile load $P=2.0$ is applied uniformly at both ends of the beam. The material properties of the beam are $E=500$ and $\nu=0.3$. The FEM solution is generated from a $100\times20\times20$ elements uniform mesh. The collocation points are the same as the grid size. The boundary conditions points are selected uniformly with the spacing of 0.005. 

To evaluate the effect of noise on the performance of PINN, we add Gaussian noise to the exact FEM solution $\boldsymbol{u}^{**}(\boldsymbol{X}_{i}^{col})$ to obtain the displacement fields $\boldsymbol{u}^*(\boldsymbol{X}_{i}^{col})$. 
The $L_2$ norm errors between $\boldsymbol{u}^*(\boldsymbol{X}_{i}^{col})$ and $\boldsymbol{u}^{**}(\boldsymbol{X}_{i}^{col})$ are $10.690\%$, $9.292\%$, and $9.914\%$ in X, Y and Z directions, respectively. 
To be more specific, the noise is sampled from the standard normal distribution (i.e. with a zero mean and a standard deviation of 1)  and then amplified to achieve the expected noise level. 
In this example, the value of $P$ is estimated based on the obtained noisy displacement field $\boldsymbol{u}^*(\boldsymbol{X}_{i}^{col})$. 

Figure~\ref{fig:3Dbeam}~(b) shows the convergence of the unknown tensile load after 20000 training epochs. The final predictions $P$ for scaled PINN, SA-PINN and PINN with uncertainty weighting are 2.0034 (error: $0.17\%$), 2.0034 (error: $0.17\%$) and 2.0022 (error: $0.11\%$), respectively. All three methods have very similar accuracy, but our approach attains the fastest convergence to the result within $1\%$ error. Figure~\ref{fig:3Dbeam}~(c) illustrates the task weights obtained from our approach. The loss term $\mathcal{L}_4$ has the smallest weight, indicating that the model focuses more on optimizing physics loss rather than the data loss corrupted by noise. In addition, we also track the mean square error between PINN predictions and the ground truth FEM solution $\boldsymbol{u}^{**}$, which reads as:
\begin{equation}
	\begin{aligned}
		MSE_{true}=\frac{1}{N_{col}}\sum_{i=1}^{N_{col}}{\left\| \boldsymbol{\bar{u}}(\boldsymbol{X}_{i}^{col};\boldsymbol{\Theta })-\boldsymbol{\bar{u}}^{**}(\boldsymbol{X}_{i}^{col}) \right\|^{2}}.
	\end{aligned}
	\label{eq:losssup}
\end{equation}

Figure~\ref{fig:3Dbeam}~(d) shows evolution of all loss terms associated with Eq.~\eqref{eq:totloss} and Eq.~\eqref{eq:losssup}. It is noteworthy that $MSE_{data}$, see Eq.~\eqref{eq:totloss}, is always fluctuating but not decreasing further after 5000 training epochs, denoting that the involvement of physics laws prevent the neural network from overfitting on noisy data. However, $MSE_{true}$ keeps dropping even to the magnitude of $10^{-7}$, which indicates that PINN method can identify the correct solution from noisy data (even those with an error of around $10\%$) and is robust to Gaussian noise.

\section{Model transfer for inverse analyses across computational domains}
\label{sec:modeltransfer}
In this section, the proposed approach is carried out on two examples to explore its possibility of inverse analyses across various computational domains under different loading conditions. Section~\ref{sec:plate4224} shows the potential of transfer learning for analyzing structural elements with geometric scaling. In Section~\ref{sec:tunnel}, we demonstrate an application of our approach to a practical inverse problem in tunnel engineering: prediction of external loading distributions of tunnel rings based on a limited number of displacement monitoring points. The performance and efficiency of transfer learning are also discussed.

\subsection{Learning the load acting on plates with geometric scaling}
\label{sec:plate4224}

\begin{figure}[!b]
	\centering
	\includegraphics{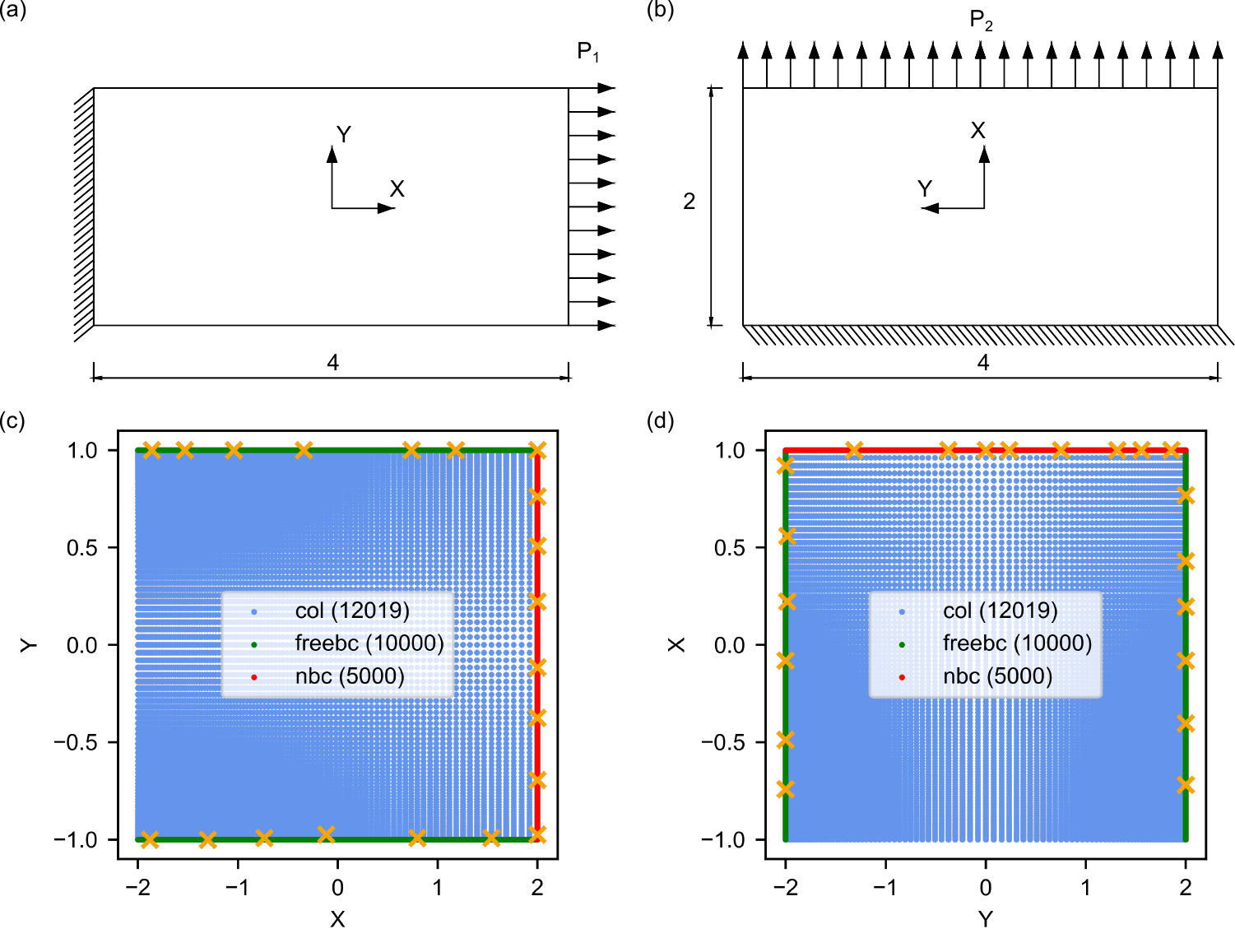}
	\caption{PINN model for prediction of the rectangular plates subjected to line loads: (a) geometry and loading condition of plate $4\times2$, (b) geometry and loading condition of plate $2\times4$, (c) training points distribution of plate $4\times2$ and (b) training points distribution of plate $2\times4$ (Observational data points are marked with orange crosses.)}
	\label{fig:plate4224_geo}
\end{figure}

Following the Algorithm~\ref{alg:algo}, we have solved a loading scenario and saved the pre-trained model at the offline stage (i.e. the square hyperelastic plate benchmark in Section~\ref{sec:plate22}). Therefore, in this section, we focus on transferring the pre-trained model to some plates with geometric scaling. As shown in Fig.~\ref{fig:plate4224_geo}~(a) and (b), two more examples, i.e. plate $4\times2$ and plate $2\times4$, are introduced, which can be regarded as the square plate benchmark after scaling transformation. The material properties and loading conditions remain consistent with the square plate, while the values of loads are changed to $P_1=P_2=15$. 

\begin{figure}[!b]
	\centering
	\includegraphics{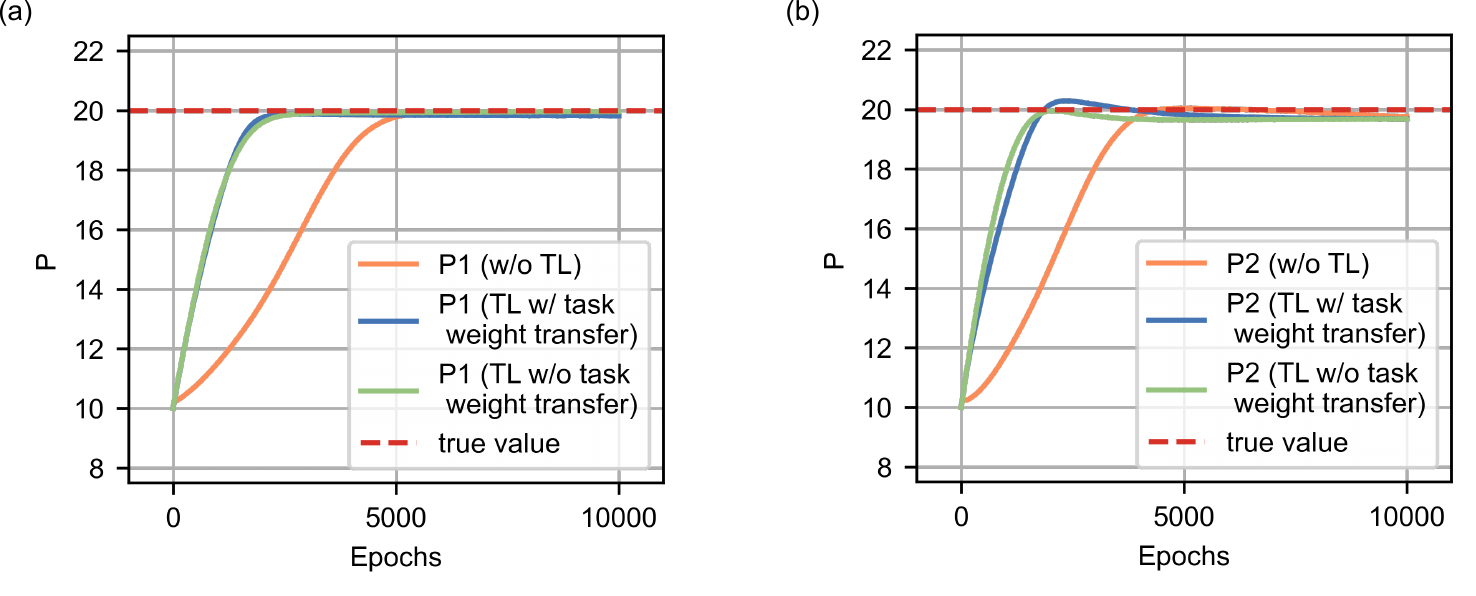}
	\caption{Results of predicted loads with (w/) and without (w/o) transfer learning (TL) after 10000 epochs for: (a) plate $4\times2$ and (b) plate $2\times4$ (Only results obtained by freezing 3 layers are plotted here.)}
	\label{plate4224_load}
\end{figure}

\begin{table}[!b]
	\centering
	\caption{A comparison of the results obtained from PINNs with and without transfer learning}
	\begin{threeparttable}
		\begin{tabular}{lcccc}
			\hline
			& \multicolumn{2}{c}{Required epoch\tnote{1}} & \multicolumn{2}{c}{Predicted load\tnote2} \\
			& $4\times2$                    & $2\times4$                   & $4\times2$                    & $2\times4$                   \\ \hline
			Without transfer learning (TL)                & 5300                          & 4100                        & 19.874                        & 19.602                       \\
			TL w/o task weight transfer (freeze 3 layers) & 2400                          & 1800                         & 19.828                        & 19.737                       \\
			TL with task weight transfer (freeze 2 layers) & 2100                          & 2700                         & 19.798                        & 19.758                       \\
			TL with task weight transfer (freeze 3 layers) & 2200                          & 1600                         & 19.807                        & 19.743                       \\
			TL with task weight transfer (freeze 4 layers) & 2300                          & 6900                        & 19.817                        & 19.824                       \\ \hline
		\end{tabular}
	\end{threeparttable}
	\begin{tablenotes}
		\footnotesize
		\item{1} The epoch here refers to the first epoch to converge to the results with $<1\%$ error.
		\item{2} The load here is obtained from the PINN after 20000 epochs.
	\end{tablenotes}
	\label{tab:platetransfer}
\end{table}

The training points distributions for two rectangular plates are shown in Fig.~\ref{fig:plate4224_geo}~(c) and~(d). The FEM nodes with local refinements at the clamped side are used as collocation points while the boundary conditions are trained with 5000 training points per edge. It is worth noting that PINNs are capable of dealing with physical systems with data collected from a limited number of sensors \cite{Karniadakis}. Therefore, the data points $\{ \boldsymbol{X}_{i}^{data} \}$ used for the inverse analysis in this section are only 20 for each case, which are randomly selected and marked with orange crosses (see Fig.~\ref{fig:plate4224_geo}).
Regarding the online stage in Algorithm~\ref{alg:algo}, we firstly froze the parameters in the first three layers of the neural network obtained from Section~\ref{sec:plate22}. Then the remaining parameters are fine-tuned with a learning rate of 0.0005 to realize the transfer learning for plates $4\times2$ and $2\times4$. 

Figure~\ref{plate4224_load} shows the results of the predicted line loads after 10000 training epochs. All the models are trained with 20000 epochs but only the results from first 10000 epochs are exhibited here for a better comparison. The results obtained from PINNs with and without transfer learning is presented in Table~\ref{tab:platetransfer}. It can be seen that transfer learning can accelerate the convergence of the model without losing accuracy. Freezing 2 or 3 layers can both generate good predictions, while keeping the initial 4 layers frozen fails to converge within the expected epochs. We attribute this to the fact that the more the neural network parameters are frozen, the lower the expressivity power is. The selection of the number of frozen layers is often a trade-off between computational efficiency (freezing more layers) and accuracy (freezing less layers). It is interesting to find that the transfer learning based PINNs still provide accurate predictions when scaling transformations are applied to the initial computational domain. Currently domain adaptation \cite{goswami_deeponetshift} are not currently taken into consideration. Nevertheless, our approach is applicable to geometrically scaled structures, which exhibits the potential for surrogate modeling of a wide range of engineering structural components (e.g. beams, plates and shells, etc.) in different sizes.

\subsection{Learning the complicated soil pressures acting on tunnels}
\label{sec:tunnel}

\begin{figure}[!b]
	\centering
	\includegraphics{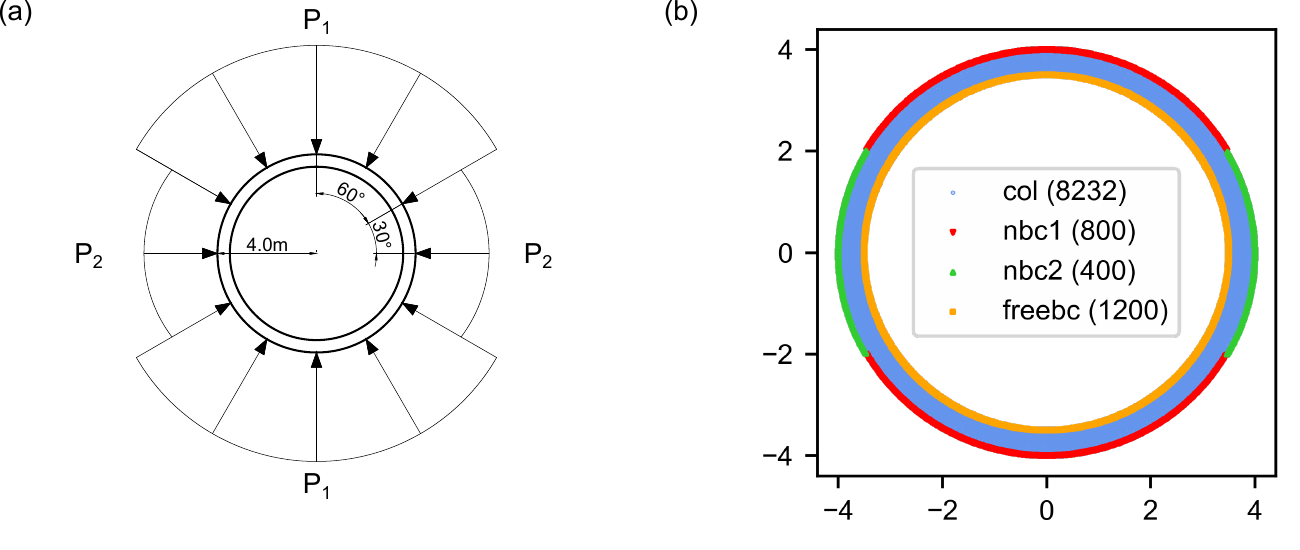}
	\caption{PINN model for prediction of loads applied on the 2D tunnel ring $2p$: (a) tunnel lining subjected to simplified vertical and lateral soil pressures ($P_1=100kN$ and $P_2=80kN$) and (b) sampling scheme of training points (Numbers of points are indicated in parentheses.)}
	\label{fig:2p_geometry}
\end{figure}

To better illustrate the complete workflow of Algorithm~\ref{alg:algo}, in this section, we manage to solve the inverse problem to quantify for the external loads acting on tunnel linings based on displacement measurements at a limited number of monitoring locations. In an actual tunneling project, it is often required to evaluate the structural behavior of the tunnel lining at multiple cross-sections. This process can be facilitated by transfer learning, in which information from the simple scenario can be used for the prediction of complicated loading scenarios at various cross-sections. The tunnel lining is considered as a 2D homogeneous isotropic ring. Given that the dimension of tunnel length is very large compared to the others, the problem can be regarded as a plane strain problem. 

First of all, the steps 1-7 at the offline stage in Algorithm~\ref{alg:algo} are carried out. Therefore, the training of PINN starts from the scratch without any prior knowledge.
As shown in Fig.~\ref{fig:2p_geometry}~(a), the tunnel ring has an outer radius of $4m$ and a thickness of $0.5m$. In light of the surrounding pressure generated by the soil medium, two uniformly distributed pressures $P_1=100kN$ and $P_2=80kN$ are directly applied on the structure, which can be considered as simplified vertical and lateral soil pressures respectively. Thus the research subject at the offline stage is referred to as case $2p$. The modulus of elasticity of concrete is $30GPa$, and the Poisson's ratio is $0.2$.  

Figure~\ref{fig:2p_geometry}~(b) depicts the training points for case $2p$. The global seed size for FEM mesh in Section~\ref{sec:tunnel} is 0.04. The FEM nodes are used as collocation points. The boundary conditions are trained with 1200 uniformly distributed training points per edge. Considering the sensor noise and environmental disturbance in the practical tunnel monitoring, some Gaussian noise are added to the exact FEM solution $\boldsymbol{u}^{**}(\boldsymbol{X}_{i}^{col})$ to obtain the displacement $\boldsymbol{u}^*(\boldsymbol{X}_{i}^{col})$ with $L_2$ errors of $11.250\%$ and $10.060\%$ in X and Y directions, respectively. 

\begin{figure}[!b]
	\centering
	\includegraphics[scale=0.93]{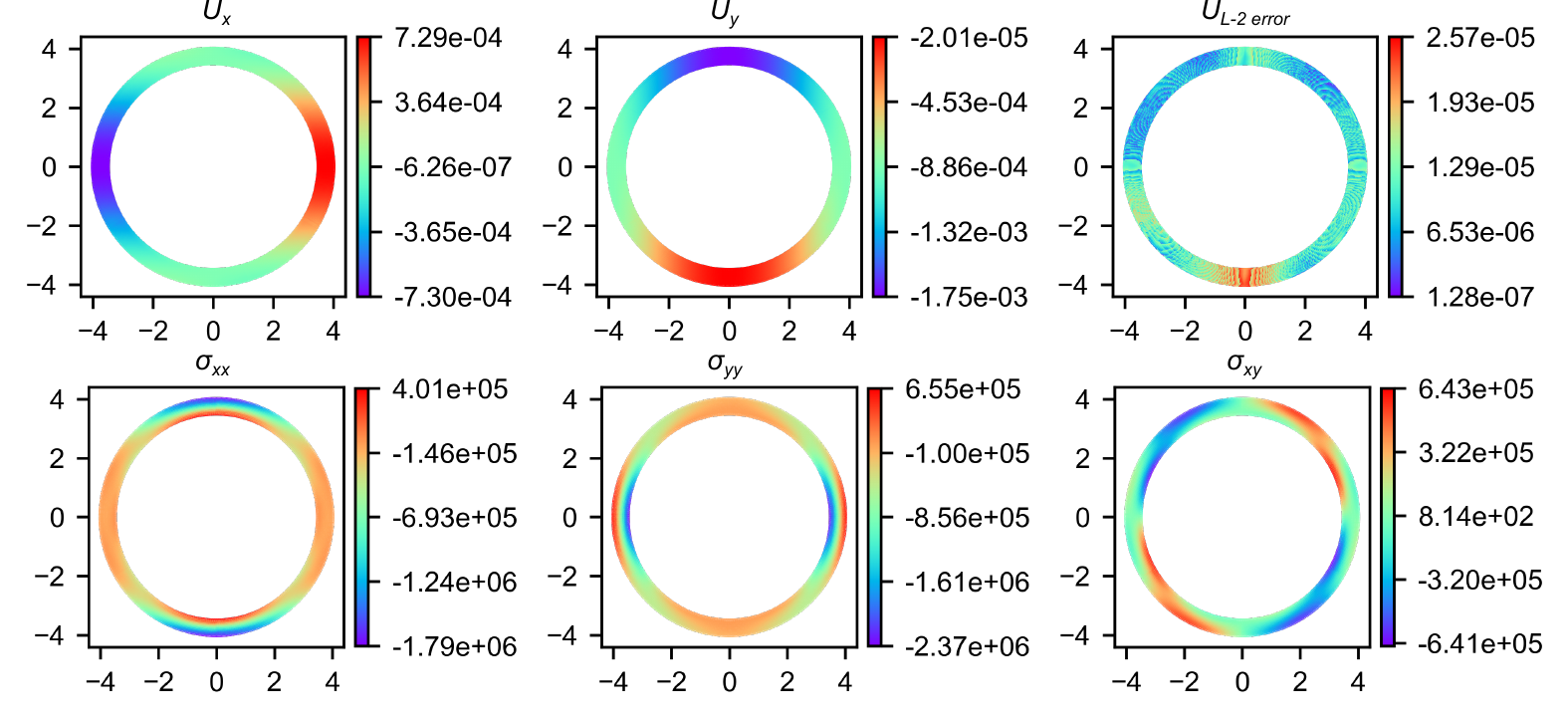}
	\caption{Resulting components of displacement vector field and stress tensor fields for case $2p$}
	\label{fig:2p_field}
\end{figure}

\begin{figure}[!b]
	\centering
	\includegraphics[scale=0.93]{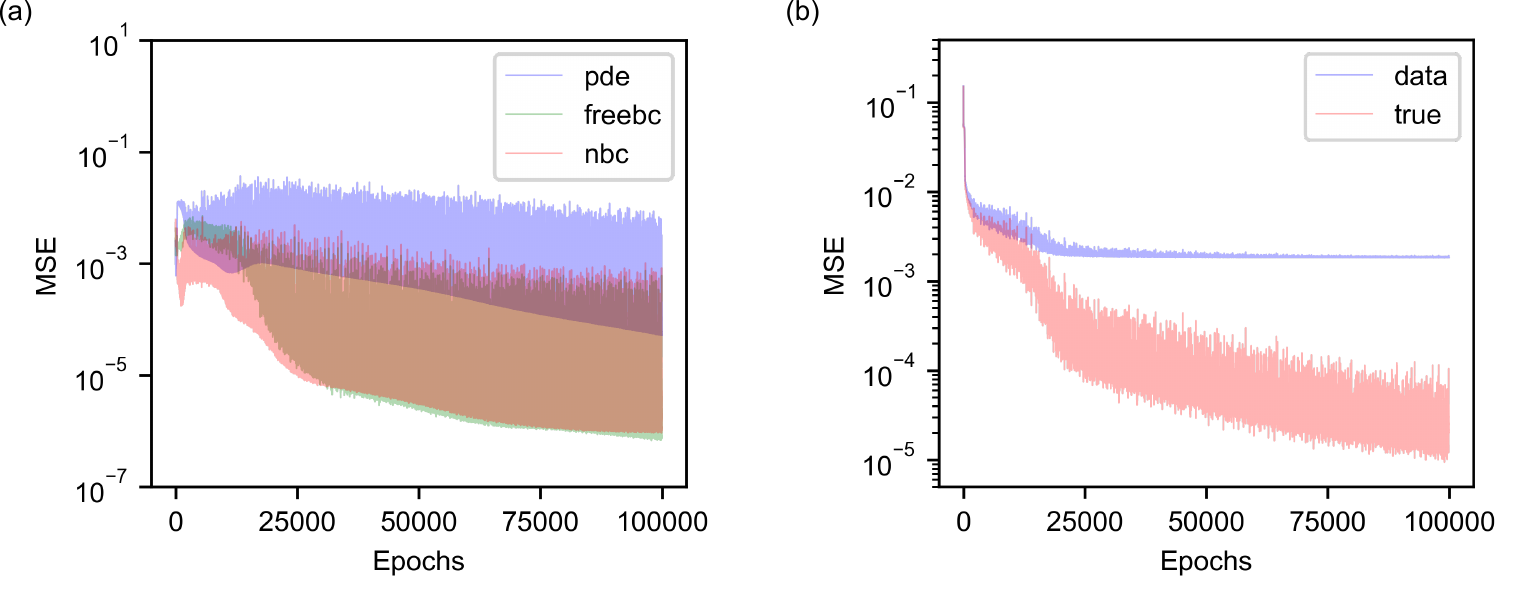}
	\caption{Convergence of loss functions for case $2p$: (a) physics loss and (b) data loss}
	\label{fig:2p_loss}
\end{figure}

\begin{figure}[!t]
	\centering
	\includegraphics{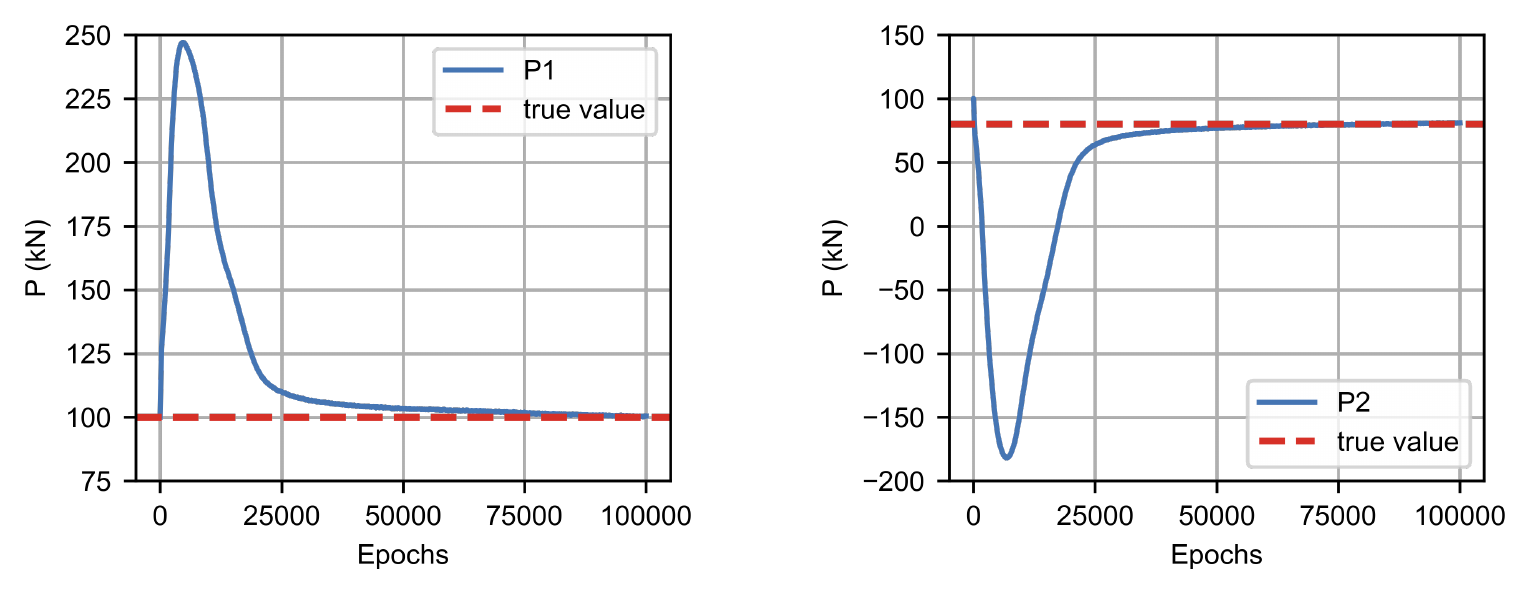}
	\caption{Convergence of external loads for case $2p$}
	\label{fig:2p_force}
\end{figure}

\begin{figure}[!t]
	\centering
	\includegraphics{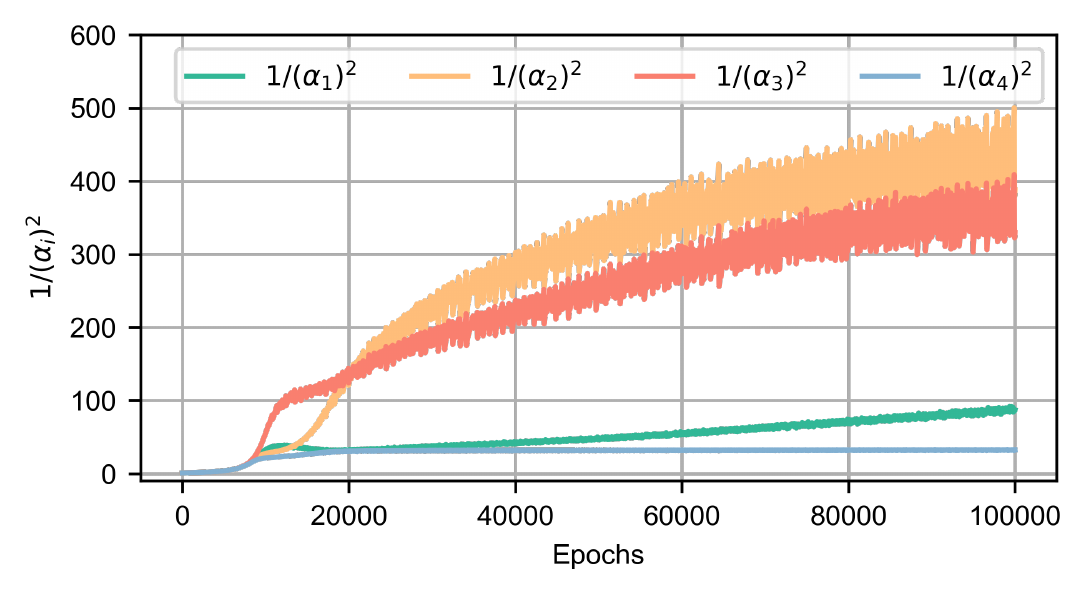}
	\caption{Results of multi-task weights for case $2p$}
	\label{fig:2p_weight}
\end{figure}

Figure~\ref{fig:2p_field} shows the displacement field as well as the stress field obtained from PINN after 100000 training epochs. The evolution of each loss term is given in Fig.~\ref{fig:2p_loss}. It can be observed that $MSE_{data}$ hardly decreases after 25000 steps but $MSE_{true}$ still keeps decreasing, which is similar to that of the 3D beam benchmark in Section~\ref{sec:3dbeam}. Figure~\ref{fig:2p_force} shows the convergence of the unknown load. The final solutions are $P_1=100.479kN$ ($0.479\%$) and $P_2=80.968kN$ ($1.210\%$), which are sufficiently accurate. Figure~\ref{fig:2p_weight} illustrates the evolution of the task weight for each loss term.

\begin{figure}[!t]
	\centering
	\includegraphics{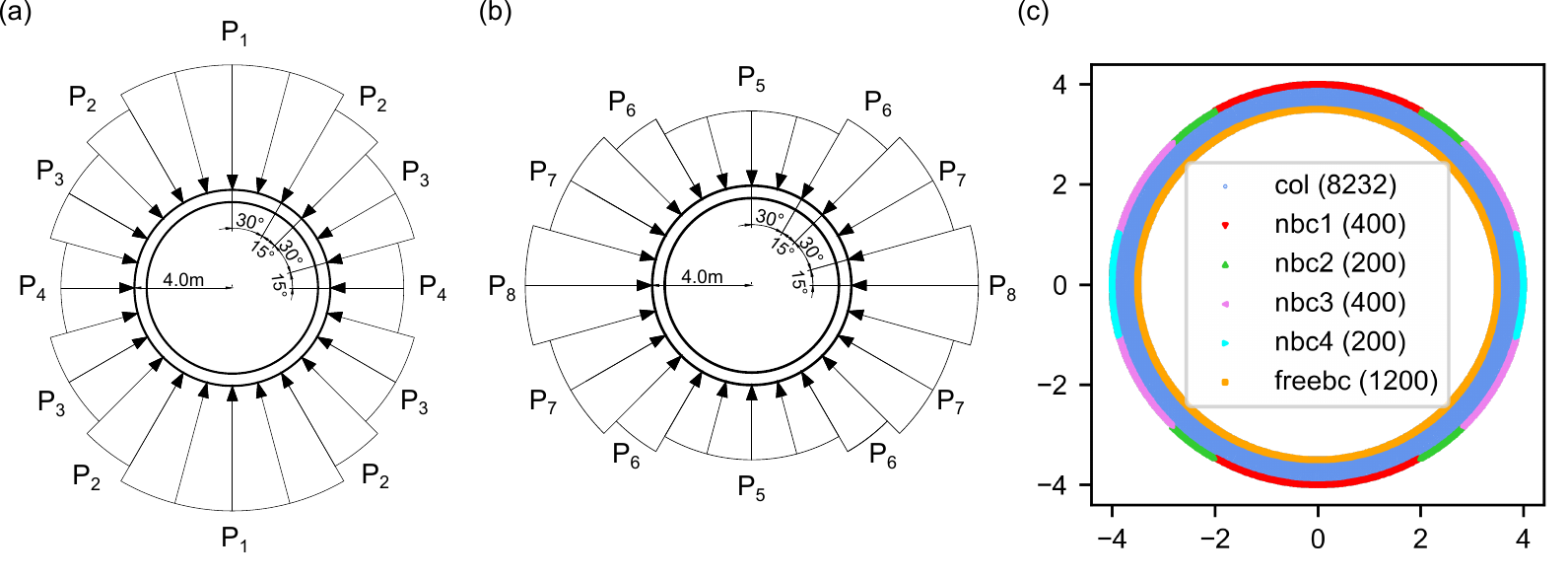}
	\caption{PINN model for prediction of loads applied on the 2D tunnel ring $4p$: (a) tunnel lining $4p\_a$ subjected to four loads ($P_1=130kN$, $P_2=120kN$, $P_3=110kN$, $P_4=100kN$), (b) tunnel lining $4p\_b$ subjected to another four loads ($P_5=100kN$, $P_6=110kN$, $P_7=120kN$, $P_8=130kN$) and (c) sampling scheme of training points (Numbers of points are indicated in parentheses.)}
	\label{fig:4p_geometry}
\end{figure}

Next, the steps 8-18 at the online stage in Algorithm~\ref{alg:algo} are carried out. The model previously trained for case $2p$ is utilized for load prediction in two more complicated loading conditions as shown in Fig.~\ref{fig:4p_geometry}~(a) and Fig.~\ref{fig:4p_geometry}~(b), where four uniformly distributed pressures are applied on the tunnel rings. Therefore, the examples at the online stage are referred to as $4p$. The structural deformation patterns are different for two loading scenarios, depending on whether the inward convergence is at the top (i.e. case $4p\_a$) or from the sides (i.e. case $4p\_b$). The geometry of the structure and the material properties remain the same as case $2p$. 

\begin{figure}[!t]
	\centering
	\includegraphics{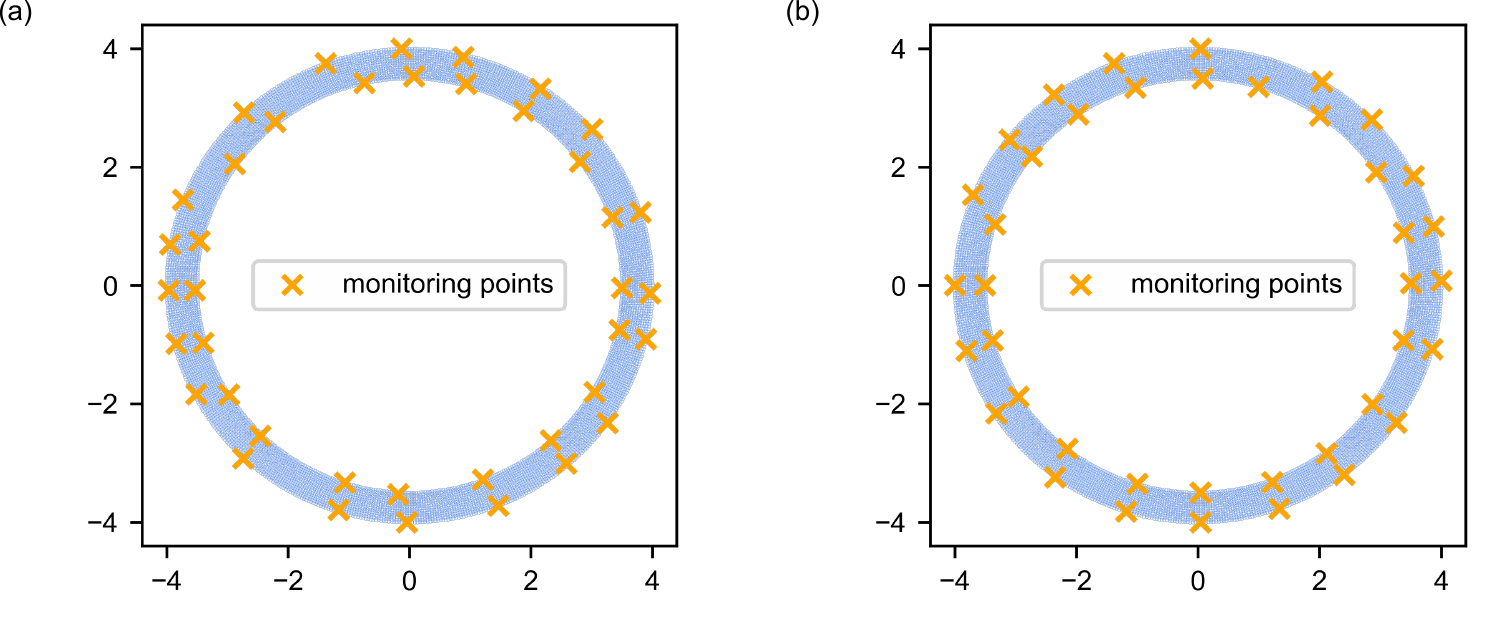}
	\caption{The layout of monitoring points for: (a) case $4p\_a$ and (b) case $4p\_b$}
	\label{fig:monitoring}
\end{figure}

Figure~\ref{fig:4p_geometry}~(c) depicts the sampling scheme of the training points for both loading cases. For a practical-oriented application in tunnel engineering, available monitoring data are assumed to be collected from a limited number of locations around the lining ring, so 40 points are randomly chosen as monitoring points for each case (see Fig.~\ref{fig:monitoring}). Some Gaussian noise are added to the exact FEM solution $\boldsymbol{u}^{**}(\boldsymbol{X}_{i}^{data})$ to obtain the displacement fields $\boldsymbol{u}^*(\boldsymbol{X}_{i}^{data})$ with $L_2$ errors of $8.015\%$ and $6.129\%$ in X and Y directions for case $4p\_a$, while $L_2$ errors in X and Y directions for case $4p\_b$ are $6.563\%$ and $7.007\%$, respectively.

\begin{figure}[!b]
	\centering
	\includegraphics{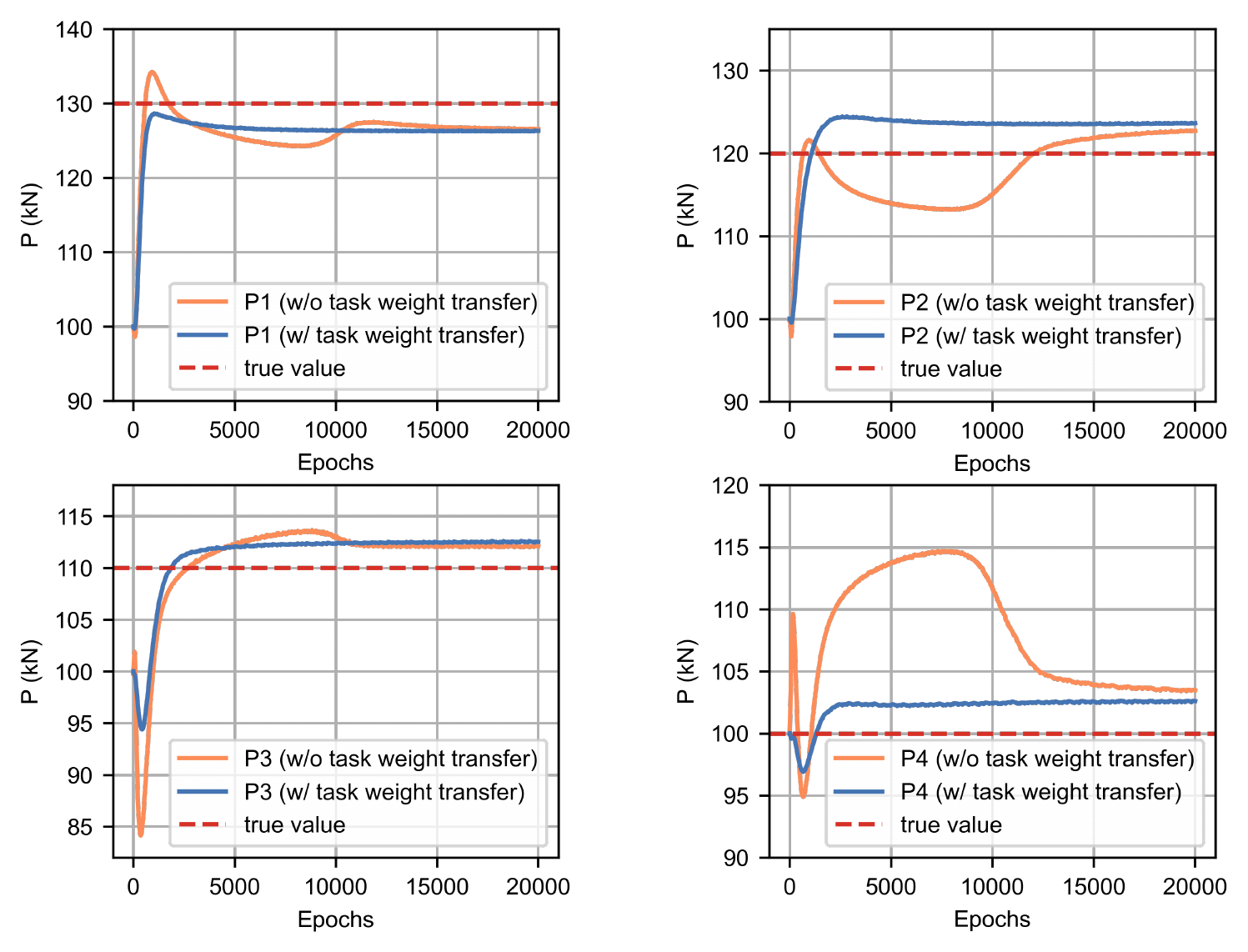}
	\caption{Convergence of external loads for case $4p\_a$ based on limited data with Gaussian noise}
	\label{fig:4pa_g_force}
\end{figure}

\begin{figure}[!b]
	\centering
	\includegraphics{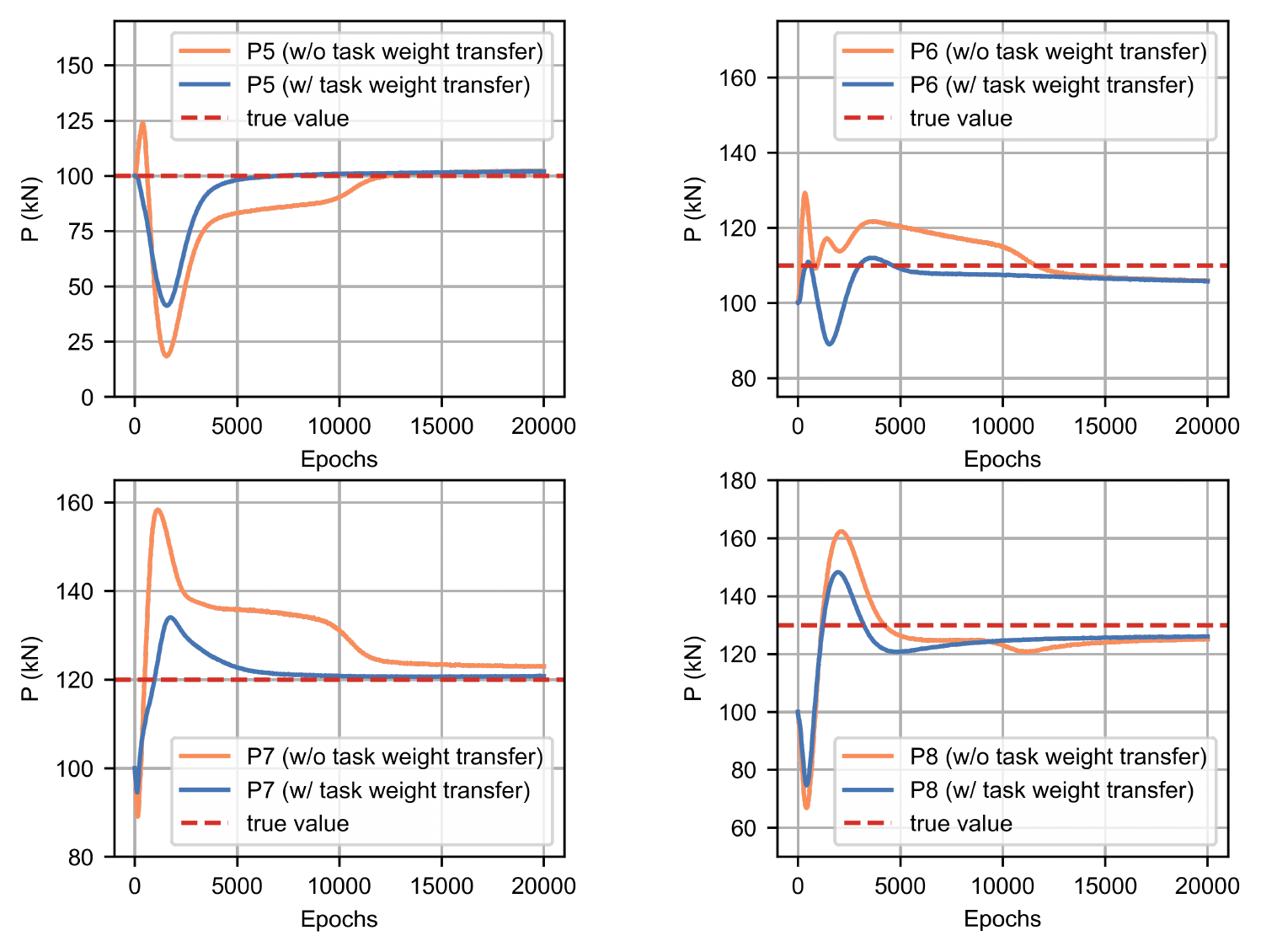}
	\caption{Convergence of external loads for case $4p\_b$ based on limited data with Gaussian noise}
	\label{fig:4pb_g_force}
\end{figure}

To better demonstrate the benefits of our multi-task learning method, the transfer learning approach without task weight transfer, i.e. all the task weights are initialized at step 13 in Algorithm~\ref{alg:algo}, is also implemented for comparison. The pre-trained PINN models are fine-tuned with 20000 training epochs. Figures~\ref{fig:4pa_g_force} and \ref{fig:4pb_g_force} show the convergence of the unknown loads. It can be seen that the PINN models converge within only 10000 steps in the case of task weight transfer, otherwise it would take 20000 steps if the task weights are initialized before fine-tuning. Tables~\ref{tab:4pa_g_force} and \ref{tab:4pb_g_force} summarize all the predicted results of unknown loads for both scenarios and the corresponding computational times.

\begin{table}[!b]
	\centering
	\caption{Final predictions of PINN for case $4p\_a$}
	\begin{tabular}{lccccc}
		\hline
		& $P_1$                                                        & $P_2$                                                        & $P_3$                                                        & $P_4$                                                        & Time      \\ \hline
		True value                                                                           & 130000                                                       & 120000                                                       & 110000                                                       & 100000                                                       & -         \\
		Without transfer learning                                                            & \begin{tabular}[c]{@{}c@{}}126846.6\\ (2.426\%)\end{tabular} & \begin{tabular}[c]{@{}c@{}}116431.6\\ (2.974\%)\end{tabular} & \begin{tabular}[c]{@{}c@{}}110513.0\\ (0.466\%)\end{tabular} & \begin{tabular}[c]{@{}c@{}}101483.4\\ (1.483\%)\end{tabular} & 50min 43s \\
		\begin{tabular}[c]{@{}l@{}}Transfer learning w/o\\ task weight transfer\end{tabular} & \begin{tabular}[c]{@{}c@{}}126562.8\\ (2.644\%)\end{tabular} & \begin{tabular}[c]{@{}c@{}}122709.7\\ (2.258\%)\end{tabular} & \begin{tabular}[c]{@{}c@{}}112091.2\\ (1.901\%)\end{tabular} & \begin{tabular}[c]{@{}c@{}}103479.3\\ (3.479\%)\end{tabular} & 17min 12s \\
		\begin{tabular}[c]{@{}l@{}}Transfer learning w/\\ task weight transfer\end{tabular}  & \begin{tabular}[c]{@{}c@{}}126375.1\\ (2.788\%)\end{tabular} & \begin{tabular}[c]{@{}c@{}}123566.9\\ (2.972\%)\end{tabular} & \begin{tabular}[c]{@{}c@{}}112358.7\\ (2.144\%)\end{tabular} & \begin{tabular}[c]{@{}c@{}}102422.9\\ (2.422\%)\end{tabular}   & 8min 22s  \\ \hline
	\end{tabular}
	\label{tab:4pa_g_force}
\end{table}

\begin{table}[!b]
	\centering
	\caption{Final predictions of PINN for case $4p\_b$}
	\begin{tabular}{lccccc}
		\hline
		& $P_5$                                                        & $P_6$                                                        & $P_7$                                                        & $P_8$                                                        & Time      \\ \hline
		True value                                                                           & 100000                                                       & 110000                                                       & 120000                                                       & 130000                                                       & -         \\
		Without transfer learning                                                            & \begin{tabular}[c]{@{}c@{}}98900.0\\ (1.099\%)\end{tabular}  & \begin{tabular}[c]{@{}c@{}}110709.1\\ (0.645\%)\end{tabular} & \begin{tabular}[c]{@{}c@{}}124559.8\\ (3.800\%)\end{tabular} & \begin{tabular}[c]{@{}c@{}}125878.6\\ (1.483\%)\end{tabular} & 52min 34s \\
		\begin{tabular}[c]{@{}l@{}}Transfer learning w/o\\ task weight transfer\end{tabular} & \begin{tabular}[c]{@{}c@{}}102180.2\\ (2.180\%)\end{tabular} & \begin{tabular}[c]{@{}c@{}}105906.2\\ (3.722\%)\end{tabular} & \begin{tabular}[c]{@{}c@{}}122966.1\\ (2.472\%)\end{tabular} & \begin{tabular}[c]{@{}c@{}}125040.2\\ (3.815\%)\end{tabular} & 17min 24s \\
		\begin{tabular}[c]{@{}l@{}}Transfer learning w/\\ task weight transfer\end{tabular}  & \begin{tabular}[c]{@{}c@{}}100881.5\\ (0.882\%)\end{tabular} & \begin{tabular}[c]{@{}c@{}}107403.3\\ (2.361\%)\end{tabular} & \begin{tabular}[c]{@{}c@{}}120751.9\\ (0.627\%)\end{tabular} & \begin{tabular}[c]{@{}c@{}}124900.2\\ (3.923\%)\end{tabular} & 8min 24s  \\ \hline
	\end{tabular}
	\label{tab:4pb_g_force}
\end{table}

\begin{figure}[!b]
	\centering
	\includegraphics{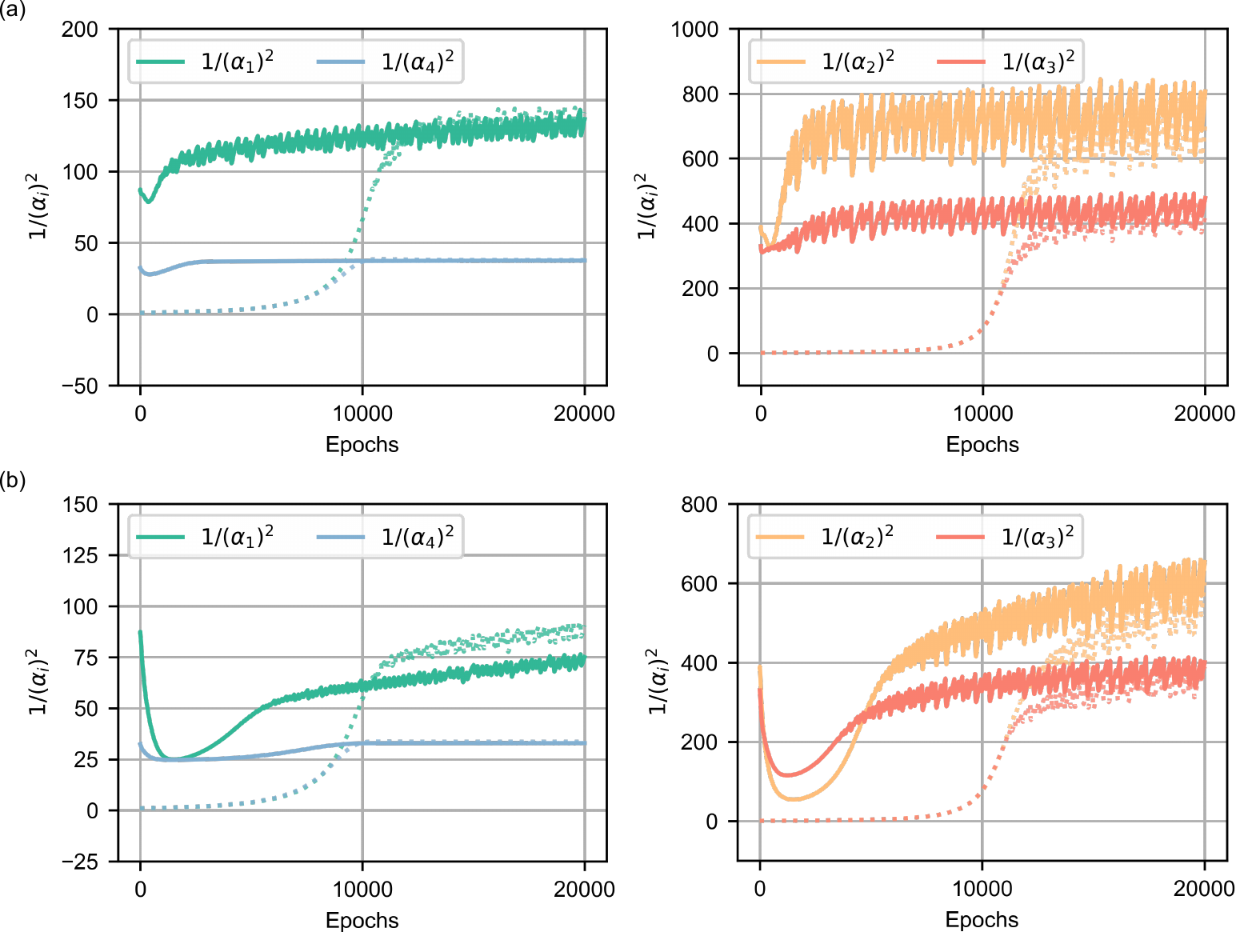}
	\caption{Results of multi-task weights for: (a) case $4p\_a$ and (b) case $4p\_b$ (Results obtained from the model without task weight transfer are shown with dotted lines.)}
	\label{fig:4p_weight}
\end{figure}

In general, the inverse analysis of predicting acting loads in both cases can be solved appropriately using PINNs with transfer learning. The average prediction errors for all four acting loads in cases $4p\_a$ and $4p\_b$ are only less than 3\%. Considering the pre-trained model in case $2p$ as the basis for transfer learning to the prediction models in both investigated cases $4p\_a$ and $4p\_b$, it is shown that the good prediction is achieved not only with a similar scenario to pre-trained scenario (cases $4p\_a$ and $2p$) but also with an unlike scenario (cases $4p\_b$ and $2p$), which shows the robustness of the proposed PINNs approach.
For a better comparison, we also provide the results obtained by training the neural networks without transfer learning in Tables~\ref{tab:4pa_g_force} and \ref{tab:4pb_g_force}. It can be seen that the prediction results of transfer learning and training from the beginning are approximately close in accuracy for both cases. As shown in Fig.~\ref{fig:transfer}~(b), the initial network parameter for a pre-trained model does not guarantee the absolute convergence to the global optimum after gradient descent, and it may easily fall into an adjacent local minimum. The PINNs without transfer learning can also converge to a physics-obeying local minimum instead of the global minimum~\cite{Bajaj_robustpinn:21, Krishnapriyan2021}. Therefore, the final accuracy of PINN predictions based on limited data with Gaussian noise strongly depends on the characteristics of loss landscape. Model transfer is also robust to Gaussian noise. The parameter freezing greatly reduces the training cost and retains some physical characteristics, which speeds up the training process. The required computing time in case of adopting transfer learning without task weight transfer is only around 17 minutes (see Tables~\ref{tab:4pa_g_force} and \ref{tab:4pb_g_force}). If the initial task weights at the online stage are also inherited from the pre-trained model, then the training time can be further reduced to around 8 min, which is suitable for real-time applications in practical tunnel projects. Figure~\ref{fig:4p_weight} displays the results of the trained multi-task weights for both scenarios. The transferred task weights are very close to the target values (i.e. well-trained task weights after 20000 epochs) at the beginning, which explains why the task weight transfer can significantly accelerate the training convergence.

\section{Conclusions}
\label{sec:Conc}
In this paper, the following conclusions can be drawn: 
(1) The non-dimensionalization and the utilization of homoscedastic uncertainty to weigh different learning tasks have significantly improved the performance of PINNs; 
(2) In case of setting the boundary conditions as learnable parameters, with enough observational data (even those with noise) and under the regularization of physics laws, PINN can successfully characterize the unknown loads of a structural system; 
(3) The introduced transfer learning based two-stage learning approach has been successfully applied to the inverse analysis in engineering structures. Although the offline model is trained previously in a simplified scenario, some physics features can be extracted and exploited, which guides the neural network at the online stage to accurately predict external loads. Good predictions of unknown acting loads are obtained not only in cases with scaled geometries but also in cases with more complicated loading scenarios compared to the boundary conditions in the pre-trained scenario. Furthermore, the requirements for observation data volume and the computational cost are both significantly reduced.

We summarize the merits of our multi-task learning approach as follows: (1) Mathematical interpretability: The task weights reflect the homoscedastic uncertainty inherent to the multiple loss terms of PINNs; (2) High efficiency: Although this method is not the most accurate, it can provide good performance with fast convergence by introducing only a small number of additional learnable parameters (consistent with the number of loss terms). No extra complex gradient calculations/operations are required; (3) Transferability: As demonstrated in~\cite{mcclenny2020self}, the adopted multi-task learning approach in our paper is in agreement with the neural network philosophy of self-adaptation, where the task weights for loss terms are regarded as parts of networks and updated by gradient descent side-by-side with the network parameters. Therefore, the saved task weights from the pre-trained model at the offline stage can be directly loaded at the online stage to boost the convergence of PINNs. 

While numerical results presented in this paper how that our approach is promising for the inverse analysis of engineering structures, some extensions to this work can be addressed in future studies. With regard to the types of loads, in this paper, we assume that the external loads are uniformly distributed on several specific regions, which means not all the boundary types (e.g. concentration loads) in engineering practice can be taken into account. When the load is applied to only a partial part of the surface, localized sharp gradients may occur in the stress components, resulting in the performance degradation of PINNs. To increase both efficiency and accuracy of the PINNs under such circumstances, the advanced network architecture proposed in~\cite{REZAEI2022115616} could be adopted as a remedy. In their new design, only first-order derivatives of the output variables with respect to the input parameters are required, and separate networks are considered for each of the output variables. Using separate neural networks not only reduces the computational cost, but also allows to specify different scaling factors for the displacement outputs in X and Y directions. Moreover, some newly proposed neural operators \cite{Lu:deeponet, Li2020, PI-deeponet} show strong inductive bias, which may help further enrich the descriptions of boundary conditions.

\section{Acknowledgment}
The first author acknowledges the support from the China Scholarship Council (CSC). The authors also gratefully acknowledge financial support by the National Natural Science Foundation of China under Grant 51808336 and the Key Project of High-speed Rail Joint Fund of National Natural Science Foundation of China under Grant U1934210. We would like to thank reviewers for taking the time and effort to review the manuscript. We sincerely appreciate all valuable comments and suggestions, which helped us to improve the quality of the manuscript.



\clearpage
\bibliographystyle{ieeetr}
\bibliography{Reference}







\end{document}